# *Stock Market Visualization*

## Zura Kakushadze[§†1] and Willie Yu[¶2]

[§] *Quantigic® Solutions LLC,[3] 1127 High Ridge Road, #135, Stamford, CT 06905*

[†] *Free University of Tbilisi, Business School & School of Physics*
*240, David Agmashenebeli Alley, Tbilisi, 0159, Georgia*

[¶] *Centre for Computational Biology, Duke-NUS Medical School*
*8 College Road, Singapore 169857*

December 27, 2017

## Abstract

We provide complete source code for a front-end GUI and its back-end counterpart for a stock market visualization tool. It is built based on the "functional visualization" concept we discuss, whereby functionality is not sacrificed for fancy graphics. The GUI, among other things, displays a color-coded signal (computed by the back-end code) based on how "out-of-whack" each stock is trading compared with its peers ("mean-reversion"), and the most sizable changes in the signal ("momentum"). The GUI also allows to efficiently filter/tier stocks by various parameters (e.g., sector, exchange, signal, liquidity, market cap) and functionally display them. The tool can be run as a web-based or local application.

---

[1] Zura Kakushadze, Ph.D., is the CEO and a Co-Founder of Quantigic® Solutions LLC and a Full Professor in the Business School and the School of Physics at Free University of Tbilisi. Email: zura@quantigic.com

[2] Willie Yu, Ph.D., is a Research Fellow at Duke-NUS Medical School. Email: willie.yu@duke-nus.edu.sg

# 1. Introduction

For human traders[4] it is a natural desire to visualize the stock market. Over the years, many stock market "heat maps" have cropped up on the web. Usually, they come and go. One of the key reasons appears to be that many of these efforts heavily focus on the graphics and miss the elephant in the room, which is functionality. The main challenge with stock market visualization tools is that the number of stocks is large (in thousands), far larger than what a (typical) human can meaningfully digest. Exacerbate that with convoluted (albeit fancy) graphics and it becomes virtually impossible to follow such a "heat map". It is simply too much information to process.

In light of this, here we discuss the concept of *functional visualization* for the stock market.[5] The idea is simple: display a large number of stock tickers on a screen without sacrificing key functionality. A few simple observations aid in this process. First, visualization must be confined to a *2-dimensional* plane: planimetry is simpler than stereometry. 3D "bubbles" and other non-planar objects are too convoluted in this context. Second, consider this: driving in Manhattan is much easier[6] than in many other cities because Manhattan (for most part) is a *square grid*. It is the 21st century and we still use Microsoft Excel, precisely because it is a square grid and data is a matrix, which allows to quickly and easily visualize it. Third, the traditional way of displaying stocks that trade up in green and those that trade down in red has limited utility. The direction on its own means little: a stock can be up because the entire market is up, but it might be down compared with its peers (e.g., its sector, industry or sub-industry). Even if this is accounted for, just "up" and "down" is insufficient as the magnitude of the price movement is also important. Displaying a relative price change as a percentage – which is numeric information – is i) difficult to follow across a large number of stocks, and ii) is not all that informative for visual perception.

What is more informative is to display how many *standard deviations* (or a similar measure) "out-of-whack" a stock is trading compared with its peers. For most stocks this number – which we refer to as a "signal" – is below 5 (outliers notwithstanding). So, it can be *color-coded* using several easy-to-distinguish colors. This is the approach we pursue here: we display tickers on a square grid with a color-coded signal. Using a square grid makes it easy to filter and tier tickers based on various informative parameters (e.g., sector, exchange, signal, liquidity, market cap). In Appendices A & B we provide complete source code for our stock market visualization tool (see Appendix C for important legalese), which we describe in detail in the remainder hereof.

---

[4] As opposed to fully automated black boxes that only require data.

[5] For prior works on stock market visualization and related literature, see, e.g., [Chang *et al*, 2007], [Csallner *et al*, 2003], [Dao *et al*, 2008], [Deboeck, 1997a,b], [Dwyer & Eades, 2002], [Eklund *et al*, 2003], [Huang *et al*, 2009], [Ingle & Deshmukh, 2017], [Jungmeister & Turo, 1992], [Keim *et al*, 2006], [Korczak & Łuszczyk, 2011], [Lin *et al*, 2005], [Ma, 2009], [Marghescu, 2007], [Novikova & Kotenko, 2014], [NYT, 2011], [Parrish, 2000], [Rehan *et al*, 2013], [Roberts, 2004], [Schreck *et al*, 2007] , [SEC, 2014], [Šimunić, 2003], [Tekušová & Kohlhammer, 2007], [Vande Moere & Lau, 2007], [Wang & Han, 2015], [Wanner *et al*, 2016], [Wattenberg, 1999], [Ziegler *et al*, 2010].

[6] Apart from aggressive taxi drivers, who are still rather innocuous compared with drivers in some other big cities.

## 2. Stock Market Visualization Map (VMap)

### 2.1. What It Does

VMap is designed to be a simple and practical (web-based) stock market visualization tool. It displays stock tickers on a square grid based on a color-coded Signal. This allows the end-user to simply and efficiently visualize how "out-of-whack" each ticker is trading compared with its industry peers (see below). VMap also allows to quickly and easily filter and/or tier tickers based on Sectors (which it refers to as Clusters), Exchanges, Liquidity and Market Capitalization.

The value of the Signal essentially represents the standard deviation[7] of the stock price movement (mathematically, its return) as compared with its peers and is computed based on a quantitative methodology we describe below. The Signal takes positive, negative and null values, not to be confused with or construed as "buy", "sell" or "hold" signals. To visualize the Signal, its absolute value is color-coded using a simple color scheme. See Table 1, Figures 1 & 2.

### 2.2. How It Works: Tiering

Stock tickers can be tiered by one of the four parameters: Sectors (Clusters), Exchanges, Liquidity, and Market Capitalization. Clusters are the same as stock Sectors and are summarized in Table 2. Exchanges are AMEX (A), NYSE (N) and NASDAQ (Q). Liquidity is based on average daily dollar volume (e.g., over the last 3 months). Tickers can be tiered by these parameters either by rows, columns, or both. So, tickers can be tiered by up to two parameters at a time. By placing a parameter on either rows or columns, tickers are split into the corresponding tiers based on that parameter. Tiering works differently for discrete and non-discrete parameters.

#### 2.2.1. Tiering: Discrete Parameters

Clusters and Exchanges are discrete parameters as they have pre-defined numbers of tiers: There are 10 Clusters (see Table 2) and 3 Exchanges (see above), each with its own pre-defined set of tickers. When two parameters are selected for tiering, tickers are displayed as blocks (corresponding to the tiers) on a square grid. VMap displays Clusters and Exchanges tiers in alphabetical ascending order, from top to bottom for rows and from left to right for columns.

#### 2.2.2. Tiering: Non-Discrete Parameters

Liquidity and Market Capitalization are non-discrete parameters and do not have pre-defined tiers. Tickers can be tiered based on these parameters by using between 2 to 10 tiers. Thus, if tiering is chosen for Liquidity, a tier number selection box appears where the Liquidity dual-slider (see below) is located on the front-end GUI (Graphical User Interface). Similarly, if tiering is chosen for Market Cap, a tier number selection box appears where the Market Cap dual-slider (see below) is located on the front-end GUI. Tiers for non-discrete parameters are generated by grouping tickers into the specified number of buckets (tiers) with approximately the same number of tickers per bucket. I.e., if $p$ tiers are chosen for a given parameter, tickers are (approximately) grouped into $p$ quantiles according to the values of that parameter. For example, if $p = 2$ tiers are chosen for Liquidity, tickers are split at the median for Liquidity – the first tier will contain the bottom 50% of tickers by Liquidity, and the second tier will contain the top 50% of tickers by Liquidity. VMap displays Liquidity and Market Capitalization tiers in the ascending order (numerically), from top to bottom for rows and from left to right for columns.

---

[7] Albeit it is computed using MAD (mean absolute deviation) – see below.

### 2.3. How It Works: Filtering

Stock tickers can be filtered based on the same parameters as in tiering. In addition, tickers can be filtered by the Signal value. Filtering works somewhat differently for various parameters.

### 2.3.1. Filtering: Discrete Parameters

Clusters can be filtered by selecting/deselecting checkboxes in the "Clusters" panel. At least one checkbox must be selected; when there is only one selected checkbox, it will be disabled. To re-enable such a checkbox, first another checkbox must be selected in the Clusters panel.

Filtering by Clusters will apply whether or not Clusters are used for tiering. If Clusters are used for tiering, deselected Clusters will be removed from the display altogether. If Clusters are not used for tiering, tickers belonging to the deselected Clusters will not be displayed.

Filtering for the Exchanges parameter is similar to filtering for the Clusters parameter.

### 2.3.2. Filtering: Liquidity and Market Capitalization

Filtering for the Liquidity parameter is done using the Liquidity dual-slider, which allows to select a range (i.e., an interval) of values. The minimum value of the Liquidity dual-slider is always zero, and the maximum value is always greater than the largest value of Liquidity across all tickers. The dual-slider scaling is pseudo-logarithmic: in the 0 to 10M section of the dual-slider each increment equals 1m; in the 10M to 100M section each increment equals 10M; etc. The Liquidity parameter can be tiered or filtered, but cannot be tiered and filtered at once.

Filtering for the Market Capitalization parameter is similar to filtering for Liquidity.

### 2.3.3. Filtering: Signal

The Signal (single-)slider scaling goes from zero to 6, and by adjusting its lower-end value one can exclude tickers with the Signal below that value. Tickers with the Signal value of 6 or higher cannot be excluded using this filtering option. Tickers for which the Signal is not available (depicted in VMAP in light gray color) are considered zero Signal tickers for filtering purposes. To exclude such tickers, the Signal slider can simply be adjusted to any desired non-zero value.

### 2.4. Flashing

The Signal is updated periodically (see below). VMap has a built-in functionality whereby it flashes up to 25 tickers with the largest absolute values of changes in their Signal values. This allows to visualize (shorter-term) "momentum" (vs. "mean-reversion" – see below) in the price movements of most active tickers. The number of flashing tickers can be adjusted using the Flashing slider. Flashing can be disabled by setting the value of the Flashing slider to zero. Some flashing tickers may not always be visible due to the screen size restrictions and/or filtering.

### 2.5. Other Features

Tickers are displayed in a square grid in VMap's main panel. Mousing over a ticker displays some pertinent information about this ticker in the lower corner on the left panel (viewing which may require scrolling down depending on the screen size). The displayed information includes the ticker symbol, Signal value (including its sign), Exchange, Cluster (Sector), Liquidity and Market Capitalization. At the top of the left panel there is a search box, which allows to look up the same information by typing in the ticker symbol and using the "Find" radio button.

### 2.6. Files Used by GUI

In Appendix A we give complete source code for the front-end GUI, whose functionality is described above. This source code was written using Adobe Flex Builder 3. Flex uses MXML and ActionScript (see Appendix A for details). Here we describe the input files `m.txt` and `s.txt` used by the GUI. The file `m.txt` is tab-separated and has 7 columns (without a header), which correspond to: a) ticker symbol; b) the numeric code for Clusters (Sectors) as defined in Table 2; c) the numeric code for Exchanges (0 = "AMEX", 1 = "NYSE", 2 = "NASDAQ"); d) Market Capitalization; e) **rank(Market Capitalization) – 1** (in the ascending order;[8] f) Liquidity; and g) **rank(Liquidity) – 1** . The ranks are pre-computed in the R code (see below) so that the GUI is spared from computing them on the fly. The file `m.txt` is generated by the R code (see below). So is the `s.txt` file. This file is a single comma/tab-separated string. It encodes information for tickers in the same order as they appear in the `m.txt` file, so it does not contain the ticker symbols. This is because the `m.txt` file has to be uploaded to the corresponding website (see Appendix A) and read by the GUI only once, typically pre-open. On the other hand, the `s.txt` file, which contains the Signal values (see below) typically is updated (frequently) intraday and must be uploaded to the corresponding website and read by the GUI accordingly (e.g., every 30 seconds). It is therefore desirable to reduce its size by including only the pertinent information.

---

[8] E.g., for Market Capitalization 100M, 10M, 20M, 1000M, **rank(Market Capitalization) – 1** would be 1, 3, 2, 0.

The first entry in the `s.txt` file is an auxiliary number, minutes-since-open (9:30 AM), rounded up. It is followed by a comma (delimiter), then the scrambled Signal (see below) for the first ticker, a tab (delimiter), then **rank(Signal) − 1** (here the rank is computed based on the unscrambled Signal), a comma (delimiter), then the scrambled Signal for the second ticker, and so on. Let the previous Signal value be **Prev.Signal**. Let **Delta = abs(Signal − Prev.Signal)**, where **abs()** is the absolute value. If for some tickers **Delta** has N/A values, such values are omitted in the following. Let **r.rank(X) = length(X) + 1 − rank(X)** be the rank of **X** in the descending order. The R code computes **r.rank(Delta) − 1** and for the tickers for which this number is less than 25 includes it in the `s.txt` file: the entry for such a ticker (which follows a delimiting comma) is: the scrambled Signal, then a tab (delimiter), then **rank(Signal) − 1**, then a tab (delimiter), then **r.rank(Delta) − 1** (followed by a delimiting comma). This is how the GUI identifies the 25 tickers (a subset of which, as determined by the Flash slider setting) should be flashed. Also, the GUI internally unscrambles the Signal, which it displays. The scrambling of the Signal is a security measure so that the `s.txt` file is essentially useless to any hacker as the Signal is scrambled.

## 3. Signal Computation

The GUI is oblivious to how the Signal is computed and in this respect can display any signal that is properly normalized (essentially, but not necessarily precisely, in the units of a standard deviation). Here, for illustrative purposes, we describe a relatively simple intraday signal, for which the "back-end" source code (written in R) is given in Appendix B. This code is essentially a call-back function, which reads an input file (see below) and outputs the `m.txt` and `s.txt` files used by the GUI (plus another file – see below). The R code does not upload or download any files. Such functionality is relegated to lower-level wrapper code (which can be written in R or any other suitable language/script). In particular, the intraday signal we discuss here typically would be recomputed frequently (e.g., every 30 seconds or even more frequently, in fact, it can be recomputed much more frequently) and the `s.txt` file would have to be uploaded to the corresponding website (see Appendix A) accordingly. This is what the wrapper code would do. It can also generate the input file for the R code (see below). The implementation of the wrapper code depends on the details of file transfer protocols used, etc., so it would make little to no sense to illustrate here: it is standard and straightforward. So we focus on the signal code.

### 3.1. Returns, MAD and Signal

Let $P_i, H_i, L_i, C_i$ be: the last intraday price after the today's open and before the today's close; today's intraday high; today's intraday low; yesterday's close price fully adjusted for any splits and dividends with the ex-date today. By "today" we mean the trading day on which we are computing the signal, and "yesterday" refers to the previous trading day. Also, the index $i$ labels stocks in our universe. First, we compute the following returns:

$$R_i = \ln\left(\frac{L_i}{X_i}\right) \tag{1}$$

The quantity $X_i$ can be computed in a number of ways. E.g., it can be set to yesterday's close $C_i$ (or today's open). We will use the following definition:

$$X_i = (1 - t)\, C_i \,+\, t\, (L_i \,+\, H_i)\,/\,2 \tag{2}$$

Here $t$ is the linear time parameter interpolating between 0 at the market open (9:30 AM) and 1 at the market close (4:00 PM). I.e., if $T$ is the actual intraday time (between 9:30 AM and 4:00 PM) measured in seconds since midnight, then

$$t = (T - 9.5 \,\times 3600)/(6.5 \,\times 3600) \tag{3}$$

So, at the open $X_i$ equals yesterday's close and as the trading day progresses it moves away from yesterday's close and gets closer to the midpoint between today's high and today's low. Note that $L_i$ and $H_i$ in Eq. (2) are understood as computed as of time $t$ (not as of today's close – intraday we do not know what today's high and low will be as of the close). It should be noted that $(L_i \,+\, H_i)\,/\,2$ is a simple (if not a layman's) way of approximating a volume-weighted average price (or VWAP), which, if available, can be used instead of $(L_i \,+\, H_i)\,/\,2$ in Eq. (2).

Now that we have our returns, we can compute an intraday signal as follows. Let us assume that we have industry classification data for our universe of tickers at the most granular level. Thus, for BICS (Bloomberg Industry Classification System) we have Sectors, Industries and Sub-industries. In this case we would take Sub-industries. Similarly, e.g., for GICS (Global Industry Classification Standard) we would take the most granular level (also called Sub-industries), for SIC (Standard Industrial Classification)[9] we would take Industries, etc. In the following, for the sake of definiteness, we will refer to the clusterings at this most granular level as "Industries". So, each ticker belongs to one and only one Industry. For some tickers we may not always have Industry data. For such tickers we will set the signal to N/A. We will omit such tickers from the computation below. For the remaining tickers, we will compute Industry averages as follows.

Let the index $A \,=\, 1, \dots, K$ label the $K$ Industies. Let $G(A)$ label the set of tickers belonging to the Industry labeled by $A$. Then the Industry averages are computed as follows:

$$R_A = \frac{1}{N(A)} \sum_{i \in G(A)} R_i\, w_i \tag{4}$$

Here $N(A)$ is the number of tickers in the set $G(A)$. The weights $w_i$ are mostly equal 1. However, for companies with multiple tickers (such as class-shares, etc.), one may wish to have nontrivial weights such that they add up to 1 when summed over all tickers for a given single company. This is because such tickers typically have almost identical returns and having them all contribute with identity weights to the Industry returns would amount to overweighing such companies' contributions. However, setting all weights $w_i = 1$ for simplicity is not a disaster.

---

[9] An open-source downloader for SIC data is provided in [Kakushadze and Yu, 2017].

We can now compute the stock returns relative to their Industry returns:

$$\tilde{R}_i = R_i - R_{M(i)} \tag{5}$$

Here $M(i)$ is the index of the Industry to which the ticker labeled by $i$ belongs (so $M$ is a map between the tickers and Industries: $M: \{1, \dots, N\} \rightarrow \{1, \dots, K\}$, wher $N$ is the number of tickers). Next, let $\sigma = \text{MAD}(\tilde{R}_i)$, where MAD is the mean absolute deviation. Then our signal is:

$$S_i = \tilde{R}_i / \sigma \tag{6}$$

This signal is related to intraday mean-reversion. Thus, a very crude strategy (which should not be construed as suitable for real-life trading) would be to go long stocks with negative $S_i$ and short stocks with positive $S_i$. Note: $S_i$ are normalized deviations from the Industry means.

### 3.2. Input Data for R Code

The R code uses a single file `mkt.data.txt` as an input. This is a tab-separated file with 12 columns and the following header: Ticker, Sector, Exchange, MktCap, Liquidity, Close, Last, High, Low, Weight, IndNames, Signal. Sector is in the numeric format (see Table 2). Exchange is also in the numeric format (see Subsection 2.5). MktCap = Market Capitalization. Liquidity (see above) is based on ADDV (average daily dollar volume), e.g., over the last 3 months. Close is fully adjusted yesterday's close (referred to as $C_i$ above). Last is the intraday last price (referred to as $P_i$ above). High is the intraday high price (referred to as $H_i$ above). Low is the intraday low price (referred to as $L_i$ above). Weight = $w_i$ above. IndNames are the Industry names (character strings); for missing industry names an empty string is used. Signal is the previous value of the Signal computed on the previous update cycle. If the Signal has not been computed yet (pre-open), NA is used. For numeric quantities such as MktCap, Liquidity, Close, Last, High and Low, missing values should be populated with zeros (not NAs) – the code handles this internally.

The R code outputs the files `m.txt` and `s.txt` used by the GUI. It also outputs a tab-separated file `sig.delta.txt` (without a header), with the following columns: scrambled Signal (see below), Signal, **rank(Signal) − 1**, **Delta**, **r.rank(Delta) − 1** (see above). This file can be used to store the Signal value to be placed in the `mkt.data.txt` file for the next update cycle. It can also be used for debugging. Finally, the scrambled Signal is computed internally using the R function `vm.scramble()` (see Appendix B), which multiplies the Signal by a factor computed as a convoluted function of the ticker's index (position) in the array of tickers, whose ordering is fixed by the `m.txt` file. The GUI then internally unscrambles the Signal.

## 4. Concluding Remarks

The advantage of the visualization scheme described in this paper is its simplicity and functionality, which largely stem from using a square grid for visualization as opposed to complex geometric shapes or 3D graphics. VMap allows the end-user to quickly and easily visualize where the "action" is: which tickers are trading "out-of-whack" compared with their peers ("mean-reversion"); which tickers have largest recent changes in the Signal (shorter-term "momentum"); tier and/or filter tickers based on Sectors, Exchanges, Signal, Liquidity and Market Capitalization, etc. The specific implementation discussed here illustrates this concept.

In this regard, we emphasize that this specific implementation is not carved in stone. Both the front-end GUI and back-end R code can be modified and optimized or even written in other languages. E.g., the GUI can be written in HTML (Hypertext Markup Language), such as HTML5 [W3C, 2014]. The choice of Flex in the instant implementation was largely motivated by a desire for "movie-like" viewing quality. Similarly, the R code can be rewritten in any suitable language.

As mentioned above, the GUI is oblivious to how the Signal is computed so long as it is normalized against some kind of a "standard deviation" such that the Signal has a reasonable distribution and the color-coding, which is designed to aid and simplify visualization, adequately captures the underlying dynamics in the market. The Signal in this particular implementation is computed using simple demeaning within Industries. More sophisticated versions of this mean-reversion-based idea via, e.g., weighted regression or optimization can be implemented. See, e.g., [Kakushadze, 2015]. Also, in the specific implementation discussed here, files are not necessarily optimally used. For instance, the `mkt.data.txt` file combines both intraday and historical data. The latter can be separated out and read once in the full implementation, which includes the wrapper code discussed above. Finally, here the input files `m.txt` and `s.txt` used by the GUI are assumed to be uploaded to a website. However, the GUI can be run locally.

## Appendix A: Source Code for Front-End GUI

Below we give complete source code for the front-end GUI ("Graphical User Interface"), whose functionality is described in the main text. This source code was written using Adobe Flex Builder 3. Flex uses MXML[10] to define UI ("User Interface") layout, ActionScript source code to address dynamic aspects, and outputs an SWF ("Small Web Format") application, which requires Adobe AIR or Flash Player at runtime to run it. In the source code below, the beginning of each file is marked using "`// BEGIN FILE...`", and the end of each file is marked using "`// END FILE...`". A few places (marked with comments using "`// COMMENT:...`") in the code refer to a URL ("Uniform Resource Locator") `www.{domain}.com/{directory}`, where the input files `m.txt` (see Section 2), `s.txt` (see Section 2) and `t.asp` (the content of this file is included below) are assumed to be located. As noted in the aforesaid comments, the placeholders `{domain}` and `{directory}` (or the entire URL) should be modified. The source code also uses the following PNG ("Portable Network Graphics") files (which are assumed to be located in the folder "`/VMap/assets/`"): `collapser.png`, `close.png`, `expander.png`, `leftThumbDown.png`, `leftThumbOver.png`, `leftThumbUp.png`, `rightThumbDown.png`, `rightThumbOver.png`, `rightThumbUp.png`, `scrollThumbDown.png`, `scrollThumbOver.png`, `scrollThumbUp.png`, `scrollTrack.png`. The reader can create these files or use the files provided in the `assets.zip` compressed folder in the Supplementary Materials.[11]

---

[10] This acronym has no official meaning. According to [Wikipedia, 2017], unofficially it is sometimes interpreted as "Magic eXtensible Markup Language" or "Macromedia eXtensible Markup Language". Adobe Systems acquired Macromedia Studio in 2005 . "Macromedia Studio MX" was one of Macromedia's products [PC Magazine, 2004].

```
// BEGIN FILE "/VMap/src/VMap.mxml"
<?xml version="1.0" encoding="utf-8"?>
<mx:Application xmlns:mx="http://www.adobe.com/2006/mxml" layout="absolute"
xmlns:controls="com.vynance.controls.*"
        applicationComplete="evt('application applicationComplete');"
        creationComplete="handle_application_creationComplete();"
        initialize="handle_application_initialize();"
        preinitialize="handle_application_preinitialize();"
        render="evt('application render');"
        resize="handle_application_resize();"
        show="evt('application show');"
        updateComplete="handle_application_updateComplete();"
xmlns:bubble="com.vynance.controls.bubble.*">

        <mx:Script><![CDATA[
                import com.vynance.controls.bubble.GlyphBubble;
                import mx.controls.Alert;
                ;
                import flash.sampler.getInvocationCount;
                import com.vynance.model.MapEvent;
                import com.vynance.model.Map;
                import com.vynance.utils.Signal;

                import com.vynance.app.AppManager;
                import com.vynance.controls.GlyphTicker;
                import com.vynance.controls.GlyphTickerEvent;
                import com.vynance.controls.dualSlider.DualSlider;
                import com.vynance.controls.dualSlider.DualSliderEvent;
                import com.vynance.model.ParamExchange;
                import com.vynance.model.ParamCluster;
                import com.vynance.model.ParamMarketCap;
                import com.vynance.model.ParamLiquidity;
                import com.vynance.modules.EnumClusters;
                import com.vynance.modules.EnumExchanges;
                import com.vynance.modules.EnumTierTypes;
                import com.vynance.model.Ticker;
```

---

[11] Which, for illustrative purposes only, also contains a sample `m.txt` file from January 1, 2016.

```actionscript
        import mx.controls.scrollClasses.ScrollBar;
import mx.controls.ToolTip;
import mx.events.ItemClickEvent;
        import mx.effects.easing.Bounce;
        import mx.formatters.NumberFormatter;
import mx.managers.ToolTipManager;

        private var _clusterCheckBoxes:Array;

        private var _currentColumn:int;

        private var _currentRow:int;

        private var _exchangeCheckBoxes:Array;

        private var _formatter:NumberFormatter;

        private function applyClusterFilter():void
        {
                var paramCluster:ParamCluster =
AppManager.getInstance().paramCluster;

                paramCluster.clearSelectedClusters();

                if (checkBoxCluster0.selected)
paramCluster.addSelectedCluster(EnumClusters.CLUSTER0);
                if (checkBoxCluster1.selected)
paramCluster.addSelectedCluster(EnumClusters.CLUSTER1);
                if (checkBoxCluster2.selected)
paramCluster.addSelectedCluster(EnumClusters.CLUSTER2);
                if (checkBoxCluster3.selected)
paramCluster.addSelectedCluster(EnumClusters.CLUSTER3);
                if (checkBoxCluster4.selected)
paramCluster.addSelectedCluster(EnumClusters.CLUSTER4);
                if (checkBoxCluster5.selected)
paramCluster.addSelectedCluster(EnumClusters.CLUSTER5);
                if (checkBoxCluster6.selected)
paramCluster.addSelectedCluster(EnumClusters.CLUSTER6);
                if (checkBoxCluster7.selected)
paramCluster.addSelectedCluster(EnumClusters.CLUSTER7);
                if (checkBoxCluster8.selected)
paramCluster.addSelectedCluster(EnumClusters.CLUSTER8);
                if (checkBoxCluster9.selected)
paramCluster.addSelectedCluster(EnumClusters.CLUSTER9);
        }

        private function applyExchangeFilter():void
        {
                var paramExchange:ParamExchange =
AppManager.getInstance().paramExchange;

                paramExchange.clearSelectedExchanges();

                if (checkBoxExchangeAMEX.selected)
paramExchange.addSelectedExchange(EnumExchanges.AMEX);
                if (checkBoxExchangeNSDQ.selected)
paramExchange.addSelectedExchange(EnumExchanges.NSDQ);
```

```actionscript
                    if (checkBoxExchangeNYSE.selected)
paramExchange.addSelectedExchange(EnumExchanges.NYSE);
            }

        private function downloadMap():void
        {

    AppManager.getInstance().map.addEventListener(MapEvent.END_SIGNAL_DOWNL
OAD , handle_Map_SignalDownloaded);

        AppManager.getInstance().map.addEventListener(MapEvent.MAP_DOWNLOADED,
handle_Map_MapDownloaded);

        AppManager.getInstance().map.addEventListener(MapEvent.MARKET_IS_CLOSED
, handle_Map_MarketIsClosed);

                    AppManager.getInstance().map.downloadMap();
            }

        public function evt(name:String):void
        {
        }

        private function findTicker(target:Object):void
        {
                for (var tickerIndex:int = 0; tickerIndex <
AppManager.getInstance().map.tickerCount(); tickerIndex++) {
                        if
(AppManager.getInstance().map.tickerByIndex(tickerIndex).symbol ==
textInputFind.text) {
                                var signal:String;
                                var ticker:Ticker;

                                ticker =
AppManager.getInstance().map.tickerByIndex(tickerIndex);

                                _formatter.precision = 2;
                                _formatter.useThousandsSeparator = true;

                                if (isNaN(ticker.signal))
                                    signal = "N/A";
                                else if (ticker.signal == 0)
                                    signal = "0.0";
                                else
                                    signal =
_formatter.format(ticker.signal);

                                formItemSymbolLabelPopup.text = ticker.symbol;
                                formItemSignalLabelPopup.text = signal;
                                formItemExchangeLabelPopup.text =
ParamExchange.getExchangeText(ticker.exchange);
                                formItemClusterLabelPopup.text =
ParamCluster.getClusterText(ticker.cluster);
                                formItemLiquidityLabelPopup.text =
ParamLiquidity.formatLiquidity(ticker.liquidity);
                                formItemMarketCapLabelPopup.text =
ParamMarketCap.formatMarketCap(ticker.marketCap);
```

```actionscript
                                glyphSignalIndicatorPopup.color =
Signal.getColor(ticker);

        glyphSignalIndicatorPopup.invalidateDisplayList();

                                currentState = "ExpandFind";

                                return;
                        }
                }

                        labelInvalidTicker.text = "Cannot find ticker " +
textInputFind.text;
                        currentState = "InvalidTicker";
                }

        private function handle_application_creationComplete():void
        {
                _formatter = new NumberFormatter();

                _clusterCheckBoxes = new Array();
                _clusterCheckBoxes.push(checkBoxCluster0);
                _clusterCheckBoxes.push(checkBoxCluster1);
                _clusterCheckBoxes.push(checkBoxCluster2);
                _clusterCheckBoxes.push(checkBoxCluster3);
                _clusterCheckBoxes.push(checkBoxCluster4);
                _clusterCheckBoxes.push(checkBoxCluster5);
                _clusterCheckBoxes.push(checkBoxCluster6);
                _clusterCheckBoxes.push(checkBoxCluster7);
                _clusterCheckBoxes.push(checkBoxCluster8);
                _clusterCheckBoxes.push(checkBoxCluster9);

                _exchangeCheckBoxes = new Array();
                _exchangeCheckBoxes.push(checkBoxExchangeAMEX);
                _exchangeCheckBoxes.push(checkBoxExchangeNSDQ);
                _exchangeCheckBoxes.push(checkBoxExchangeNYSE);

                this.addEventListener(DualSliderEvent.CHANGE,
handle_DualSlider_CHANGE);

                this.addEventListener(GlyphTickerEvent.ROLL_OUT,
handle_GlyphTicker_ROLL_OUT);
                this.addEventListener(GlyphTickerEvent.ROLL_OVER,
handle_GlyphTicker_ROLL_OVER);
        }

        private function handle_application_initialize():void
        {
        }

        private function handle_application_preinitialize():void
        {
        }

        private function handle_application_resize():void
        {
```

```actionscript
        }

        private function handle_application_updateComplete():void
        {
                if (!AppManager.getInstance().mapDownloaded) {
                        AppManager.getInstance().mapDownloaded = true;
                        this.callLater(downloadMap);
                }
        }

        private function handle_buttonFind_click():void
        {
                findTicker(buttonFind);
        }

        private function handle_canvasMain_resize(e:Event):void
        {
                evt("canvas resize");

                if (vBoxSettings == null) return;

                AppManager.getInstance().mainCanvasResized = true;

                vBoxSettings.height = this.height - 27 - 20;

                matrix.width = this.width - vBoxSettings.x -
vBoxSettings.width;
                matrix.height = this.height - 20;
        }

        private function
handle_checkBoxCluster_change(clusterIndex:int):void
        {
                var checkBox:CheckBox;

                updateModel();

                if
(AppManager.getInstance().paramCluster.countOfSelectedClusters == 1) {
                        for each (checkBox in _clusterCheckBoxes) {
                                checkBox.enabled = !checkBox.selected;
                        }
                }
                else {
                        for each (checkBox in _clusterCheckBoxes) {
                                checkBox.enabled = true;
                        }
                }
        }

        private function
handle_checkBoxExchange_change(exchangeIndex:int):void
        {
                var checkBox:CheckBox;

                updateModel();
```

```actionscript
                    if
(AppManager.getInstance().paramExchange.countOfSelectedExchanges == 1) {
                        for each (checkBox in _exchangeCheckBoxes) {
                            checkBox.enabled = !checkBox.selected;
                        }
                    }
                    else {
                        for each (checkBox in _exchangeCheckBoxes) {
                            checkBox.enabled = true;
                        }
                    }
                }

            private function handle_collapser_click(collapser:Collapser,
panel:Panel, height:int):void
                {
                    collapser.toggle();
                    if (collapser.collapsed) {
                        panel.height = 29;
                        collapser.toolTip = "Expand";
                    }
                    else {
                        panel.height = height;
                        collapser.toolTip = "Collapse";
                    }
                }

            private function handle_comboBoxLiquidity_change():void
                {
                    updateModel();
                }

            private function handle_comboBoxMarketCap_change():void
                {
                    updateModel();
                }

            private function
handle_DualSlider_CHANGE(event:DualSliderEvent):void
                {
                    if (DualSlider(event.target).id == "dualSliderLiquidity") {
                        AppManager.getInstance().paramLiquidity.min      =
event.min * 1000000;
                        AppManager.getInstance().paramLiquidity.max      =
event.max * 1000000;
                    }
                    else if (DualSlider(event.target).id ==
"dualSliderMarketCap") {
                        AppManager.getInstance().paramMarketCap.min      =
event.min * 1000000;
                        AppManager.getInstance().paramMarketCap.max      =
event.max * 1000000;
                    }
                    else {
                        trace("ERROR");
                    }
                    updateModel();
```

```actionscript
            }

            private function
handle_GlyphTicker_ROLL_OUT(event:GlyphTickerEvent):void
            {
                    formItemSymbolLabel.text = "";
                    formItemSignalLabel.text = "";
                    glyphSignalIndicator.visible = false;
                    formItemExchangeLabel.text = "";
                    formItemClusterLabel.text = "";
                    formItemLiquidityLabel.text = "";
                    formItemMarketCapLabel.text = "";
            }

            private function
handle_GlyphTicker_ROLL_OVER(event:GlyphTickerEvent):void
            {
                    var glyphTicker:GlyphTicker;
                    var signal:String;
                    var ticker:Ticker;
                    var tickerIndex:int;

                    glyphTicker = (event.target as GlyphTicker);
                    if (glyphTicker == null)
                          tickerIndex = GlyphBubble(event.target).tickerIndex;
                    else
                          tickerIndex = GlyphTicker(event.target).tickerIndex;
                    ticker =
AppManager.getInstance().map.tickerByIndex(tickerIndex);

                    _formatter.precision = 2;
                    _formatter.useThousandsSeparator = true;

                    if (isNaN(ticker.signal))
                          signal = "N/A";
                    else if (ticker.signal == 0)
                          signal = "0.0";
                    else
                          signal = _formatter.format(ticker.signal);

                    formItemSymbolLabel.text = ticker.symbol;
                    formItemSignalLabel.text = signal;
                    formItemExchangeLabel.text =
ParamExchange.getExchangeText(ticker.exchange);
                    formItemClusterLabel.text =
ParamCluster.getClusterText(ticker.cluster);
                    formItemLiquidityLabel.text =
ParamLiquidity.formatLiquidity(ticker.liquidity);
                    formItemMarketCapLabel.text =
ParamMarketCap.formatMarketCap(ticker.marketCap);

                    glyphSignalIndicator.color = Signal.getColor(ticker);
                    glyphSignalIndicator.invalidateDisplayList();
                    glyphSignalIndicator.visible = true;
            }

            private function handle_hSliderFlashing_change():void
```

```
                {
                        if (hSliderFlashing.value.toString() == "0")
                                panelFlashing.status = "Disabled";
                        else
                                panelFlashing.status =
hSliderFlashing.value.toString();
                        AppManager.getInstance().flashingRank =
int(hSliderFlashing.value);

                        matrix.invalidateProperties();
                        matrix.invalidateDisplayList();
                }

                private function handle_hSliderSignal_change():void
                {
                        panelSignal.status = ">=" + hSliderSignal.value.toString();
                        updateModel();
                }

                private function handle_Map_MapDownloaded(event:Event):void
                {
                        var valuesLiquidity:Array = new Array();
                        var valuesMarketCap:Array = new Array();
                        var captionsLiquidity:Array = new Array();
                        var captionsMarketCap:Array = new Array();
                        var markersLiquidity:Array = new Array();
                        var markersMarketCap:Array = new Array();

                        var value:int;
                        var maxValue:int;
                        var i:int;
                        var multiplier:int;

                        value = int(AppManager.getInstance().map.minLiquidity /
1000000);
                        maxValue = int(AppManager.getInstance().map.maxLiquidity /
1000000);
                        i = 0;
                        multiplier = 1;
                        while (true) {
                                valuesLiquidity[i] = value;
                                if (value == 0)
                                        captionsLiquidity[i] = "0";
                                else if (value < 1000)
                                        captionsLiquidity[i] = value.toString() + "m";
                                else
                                        captionsLiquidity[i] = (value /
1000).toString() + "b";

                                if (value == multiplier * 10) {
                                        markersLiquidity[i] = captionsLiquidity[i];
                                        multiplier *= 10;
                                }
                                if (value > maxValue)
                                        break;
                                i++;
                                value += multiplier;
                        }
```

```actionscript
                var dualSliderLiquidity:DualSlider = new
DualSlider(valuesLiquidity, captionsLiquidity, markersLiquidity);
                dualSliderLiquidity.height = 60;
                dualSliderLiquidity.id = "dualSliderLiquidity";
                dualSliderLiquidity.width = 165;
                dualSliderLiquidity.x =
(canvasLiquidityFilterSettings.width - 165) / 2;

        canvasLiquidityFilterSettings.addChild(dualSliderLiquidity);

                value = int(AppManager.getInstance().map.minMarketCap /
1000000);
                maxValue = int(AppManager.getInstance().map.maxMarketCap /
1000000);
                i = 0;
                multiplier = 1;
                while (true) {
                    valuesMarketCap[i] = value;
                    if (value == 0)
                        captionsMarketCap[i] = "0";
                    else if (value < 1000)
                        captionsMarketCap[i] = value.toString() + "m";
                    else
                        captionsMarketCap[i] = (value /
1000).toString() + "b";
                    if (value == multiplier * 10) {
                        markersMarketCap[i] = captionsMarketCap[i];
                        multiplier *= 10;
                    }
                    if (value > maxValue)
                        break;
                    i++;
                    value += multiplier;
                }
                var dualSliderMarketCap:DualSlider = new
DualSlider(valuesMarketCap, captionsMarketCap, markersMarketCap);
                dualSliderMarketCap.height = 60;
                dualSliderMarketCap.id = "dualSliderMarketCap";
                dualSliderMarketCap.width = 165;
                dualSliderMarketCap.x =
(canvasMarketCapFilterSettings.width - 165) / 2;

        canvasMarketCapFilterSettings.addChild(dualSliderMarketCap);

            }

            private function handle_Map_MarketIsClosed(event:Event):void
            {
                matrix.invalidateProperties();
                matrix.invalidateDisplayList();

                Alert.show("Markets are now closed. Please reload the page
when markets open.");
            }

            private function handle_Map_SignalDownloaded(event:Event):void
            {
```

```
                        handle_hSliderFlashing_change();

                        if (!canvasMain.enabled)
                                canvasMain.enabled = true;

                        if (!matrix.visible)
                                this.callLater(showMatrix);
                }

                private function handle_radioButtonGroup_itemClick():void
                {
                        const SELECTED_COLOR:String = "#e5e5e5";

                        radioButtonClustersOnColumns.visible =
!radioButtonClustersOnRows.selected;
                        radioButtonExchangesOnColumns.visible =
!radioButtonExchangesOnRows.selected;
                        radioButtonLiquidityOnColumns.visible =
!radioButtonLiquidityOnRows.selected;
                        radioButtonMarketCapOnColumns.visible =
!radioButtonMarketCapOnRows.selected;

                        radioButtonClustersOnRows.visible =
!radioButtonClustersOnColumns.selected;
                        radioButtonExchangesOnRows.visible =
!radioButtonExchangesOnColumns.selected;
                        radioButtonLiquidityOnRows.visible =
!radioButtonLiquidityOnColumns.selected;
                        radioButtonMarketCapOnRows.visible =
!radioButtonMarketCapOnColumns.selected;

                        if (radioButtonClustersOnColumns.selected ||
radioButtonClustersOnRows.selected) {
                                labelTiersClusters.setStyle("color", SELECTED_COLOR);
                                labelTiersClusters.setStyle("fontWeight", "bold");
                                if (radioButtonClustersOnColumns.selected)
                                        panelClusters.status = "On Columns";
                                else
                                        panelClusters.status = "On Rows";
                        }
                        else {
                                labelTiersClusters.setStyle("color", "#ff9000");
                                labelTiersClusters.setStyle("fontWeight", "normal");
                                panelClusters.status = "";
                        }

                        if (radioButtonExchangesOnColumns.selected ||
radioButtonExchangesOnRows.selected) {
                                labelTiersExchanges.setStyle("color",
SELECTED_COLOR);
                                labelTiersExchanges.setStyle("fontWeight", "bold");
                                if (radioButtonExchangesOnColumns.selected)
                                        panelExchanges.status = "On Columns";
                                else
                                        panelExchanges.status = "On Rows";
                        }
                        else {
```

```actionscript
                    labelTiersExchanges.setStyle("color", "#ff9000");
                    labelTiersExchanges.setStyle("fontWeight", "normal");
                    panelExchanges.status = "";
            }

            if (radioButtonLiquidityOnColumns.selected ||
radioButtonLiquidityOnRows.selected) {
                    labelTiersLiquidity.setStyle("color",
SELECTED_COLOR);
                    labelTiersLiquidity.setStyle("fontWeight", "bold");
                    canvasLiquidityFilterSettings.visible = false;
                    canvasLiquidityTierSettings.visible = true;
                    canvasLiquidityTierSettings.y = 0;
                    if (radioButtonLiquidityOnColumns.selected)
                            panelLiquidity.status = "On Columns";
                    else
                            panelLiquidity.status = "On Rows";
            }
            else {
                    labelTiersLiquidity.setStyle("color", "#ff9000");
                    labelTiersLiquidity.setStyle("fontWeight", "normal");
                    canvasLiquidityFilterSettings.visible = true;
                    canvasLiquidityFilterSettings.y = 0;
                    canvasLiquidityTierSettings.visible = false;
                    panelLiquidity.status = "";
            }

            if (radioButtonMarketCapOnColumns.selected ||
radioButtonMarketCapOnRows.selected) {
                    labelTiersMarketCap.setStyle("color",
SELECTED_COLOR);
                    labelTiersMarketCap.setStyle("fontWeight", "bold");
                    canvasMarketCapFilterSettings.visible = false;
                    canvasMarketCapTierSettings.visible = true;
                    canvasMarketCapTierSettings.y = 0;
                    if (radioButtonMarketCapOnColumns.selected)
                            panelMarketCap.status = "On Columns";
                    else
                            panelMarketCap.status = "On Rows";
            }
            else {
                    labelTiersMarketCap.setStyle("color", "#ff9000");
                    labelTiersMarketCap.setStyle("fontWeight", "normal");
                    canvasMarketCapFilterSettings.visible = true;
                    canvasMarketCapFilterSettings.y = 0;
                    canvasMarketCapTierSettings.visible = false;
                    panelMarketCap.status = "";
            }

            updateModel();
        }

        private function
handle_textInputFind_keyDown(event:KeyboardEvent):void
        {
            if (event.keyCode == 13)
                    findTicker(textInputFind);
```

```actionscript
                    if (event.keyCode == 27)
                            currentState = null;
            }

            private function showMatrix():void
            {
                    trace("showMatrix");
                    matrix.visible = true;
            }

            private function updateModel():void
            {
                    if (radioButtonClustersOnColumns.selected) {
                            AppManager.getInstance().paramCluster.tierType =
EnumTierTypes.Columns;
                    }
                    else if (radioButtonClustersOnRows.selected) {
                            AppManager.getInstance().paramCluster.tierType =
EnumTierTypes.Rows;
                    }
                    else {
                            AppManager.getInstance().paramCluster.tierType =
EnumTierTypes.None;
                    }
                    applyClusterFilter();

                    if (radioButtonExchangesOnColumns.selected) {
                            AppManager.getInstance().paramExchange.tierType =
EnumTierTypes.Columns;
                    }
                    else if (radioButtonExchangesOnRows.selected) {
                            AppManager.getInstance().paramExchange.tierType =
EnumTierTypes.Rows;
                    }
                    else {
                            AppManager.getInstance().paramExchange.tierType =
EnumTierTypes.None;
                    }
                    applyExchangeFilter();

                    if (radioButtonLiquidityOnColumns.selected) {
                            AppManager.getInstance().paramLiquidity.tierType =
EnumTierTypes.Columns;
                            AppManager.getInstance().paramLiquidity.numberOfTiers
= int(comboBoxLiquidity.value);
                    }
                    else if (radioButtonLiquidityOnRows.selected) {
                            AppManager.getInstance().paramLiquidity.tierType =
EnumTierTypes.Rows;
                            AppManager.getInstance().paramLiquidity.numberOfTiers
= int(comboBoxLiquidity.value);
                    }
                    else {
                            AppManager.getInstance().paramLiquidity.tierType =
EnumTierTypes.None;
                    }
```

```
                    if (radioButtonMarketCapOnColumns.selected) {
                            AppManager.getInstance().paramMarketCap.tierType =
EnumTierTypes.Columns;
                            AppManager.getInstance().paramMarketCap.numberOfTiers
= int(comboBoxMarketCap.value);
                    }
                    else if (radioButtonMarketCapOnRows.selected) {
                            AppManager.getInstance().paramMarketCap.tierType =
EnumTierTypes.Rows;
                            AppManager.getInstance().paramMarketCap.numberOfTiers
= int(comboBoxMarketCap.value);
                    }
                    else {
                            AppManager.getInstance().paramMarketCap.tierType =
EnumTierTypes.None;
                    }

                    AppManager.getInstance().paramSignal.min =
hSliderSignal.value;

                    AppManager.getInstance().map.recalculate();

                    matrix.invalidateProperties();
                    matrix.invalidateDisplayList();

            }

    ]]></mx:Script>

    <mx:Style>

            Application {
                    backgroundColor: #121212;
                    font-size: 9;
                    paddingLeft: 0px;
                    paddingRight: 0px;
                    paddingTop: 0px;
                    paddingBottom: 0px;
            }

            CheckBox {
                    text-roll-over-color: #ffd091;
                    text-selected-color: #ffffff;
            }

            ComboBox {
                    alternating-item-colors: #cccccc, #bbbbbb;
                    color: #404040;
            }

            Form {
                    padding-bottom: 0;
                    padding-left: 0;
                    padding-right: 0;
                    padding-top: 0;
                    vertical-gap: 0;
            }
```

```
Label {
        color: #ff9000;
}

Panel {
        background-color: #404040;
        border-color: #383838;
        border-thickness-left: 5;
        border-thickness-right: 5;
        color: #ff9000;
        corner-radius: 8;
        header-height: 16;
        shadow-direction: right;
        shadow-distance: 1;
        title-style-name: "panelTitle";
}

ScrollBar {
        up-arrow-skin: ClassReference(null);
        down-arrow-skin: ClassReference(null);
        trackSkin: Embed(source="../assets/scrollTrack.png",
scaleGridLeft="1", scaleGridTop="2", scaleGridRight="2",
scaleGridBottom="3");
        thumbUpSkin: Embed(source="../assets/scrollThumbUp.png",
scaleGridLeft="1", scaleGridTop="1", scaleGridRight="3",
scaleGridBottom="3");
        thumbOverSkin:
Embed(source="../assets/scrollThumbOver.png", scaleGridLeft="1",
scaleGridTop="1", scaleGridRight="3", scaleGridBottom="3");
        thumbDownSkin:
Embed(source="../assets/scrollThumbDown.png", scaleGridLeft="1",
scaleGridTop="1", scaleGridRight="3", scaleGridBottom="3");
}

TextInput {
        borderStyle: none;
        backgroundColor: #cccccc;
        fontSize: 9;
        fontWeight: bold;
}

ToolTip {
        backgroundAlpha: 0.9;
        backgroundColor: #ffcc99;
        cornerRadius: 4;
        fontSize: 10;
}

.canvasFind {
        background-alpha: 0.95;
        background-color: #121212;
        border-color: #383838;
        border-style: solid;
        corner-radius: 8;
}
```

```
            .panelTitle {
                color: #e5e5e5;
                font-weight: bold;
            }

    </mx:Style>

    <mx:Fade alphaFrom="0.0" alphaTo="1.0" duration="2000" id="fadeIn"/>

    <mx:transitions>
     <mx:Transition>
         <mx:Parallel targets="{[canvasFind]}">
             <mx:Resize duration="1000" easingFunction="Bounce.easeOut"/>
             <mx:Sequence target="{formFind}">
                 <mx:Blur duration="250" blurYFrom="1.0" blurYTo="20.0"/>
                 <mx:Blur duration="250" blurYFrom="20.0" blurYTo="1"/>
             </mx:Sequence>
         </mx:Parallel>
     </mx:Transition>
    </mx:transitions>

    <mx:states>
        <mx:State name="ExpandFind">
            <mx:SetProperty target="{canvasFind}" name="height" value="142"/>
            <mx:SetProperty target="{formFind}" name="visible" value="true"/>
            <mx:SetProperty target="{glyphSignalIndicatorPopup}"
name="visible" value="true"/>
            <mx:SetProperty target="{labelInvalidTicker}" name="visible"
value="false"/>
            <mx:SetStyle target="{canvasFind}" name="borderColor"
value="0xff9000"/>
        </mx:State>
        <mx:State name="InvalidTicker">
            <mx:SetProperty target="{canvasFind}" name="height" value="48"/>
            <mx:SetProperty target="{formFind}" name="visible"
value="false"/>
            <mx:SetProperty target="{glyphSignalIndicatorPopup}"
name="visible" value="false"/>
            <mx:SetProperty target="{labelInvalidTicker}" name="visible"
value="true"/>
            <mx:SetStyle target="{canvasFind}" name="borderColor"
value="0xff9000"/>
        </mx:State>
    </mx:states>

    <mx:Canvas backgroundColor="#121212" enabled="false" height="100%"
horizontalScrollPolicy="off" id="canvasMain" minHeight="290"
verticalScrollPolicy="off" width="100%"
        addedToStage="evt('canvas addedToStage');"
        creationComplete="evt('canvas creationComplete');"
        initialize="evt('canvas initialize');"
        preinitialize="evt('canvas preinitialize');"
        render="evt('canvas render');"
        resize="handle_canvasMain_resize(event);"
        show="evt('canvas show');">
        <mx:VBox id="vBoxSettings" height="100%" left="20" top="36"
width="222" x="10" y="10" verticalScrollPolicy="auto">
```

```
                    <mx:Panel height="145" id="panelRowsAndColumns"
layout="absolute" title="Tiers" width="200">
                        <controls:Collapser
click="handle_collapser_click(collpserRowsAndColumns, panelRowsAndColumns,
145);" id="collpserRowsAndColumns" toolTip="Collapse" x="181" y="1"/>
                        <mx:Label text="Rows" x="92" y="2"/>
                        <mx:Label text="Columns" x="131" y="2"/>
                        <mx:Label id="labelTiersEmpty" text="Empty"
textAlign="right" width="66" x="10" y="20"/>
                        <mx:Label id="labelTiersClusters" text="Clusters"
textAlign="right" width="66" x="10" y="40"/>
                        <mx:Label id="labelTiersExchanges" text="Exchanges"
textAlign="right" width="66" x="10" y="60"/>
                        <mx:Label id="labelTiersLiquidity" text="Liquidity"
textAlign="right" width="66" x="10" y="80"/>
                        <mx:Label id="labelTiersMarketCap" text="Market Cap"
textAlign="right" width="66" x="10" y="100"/>
                        <mx:Canvas x="100" y="20">
                            <mx:RadioButtonGroup id="radioButtonGroupRows"
itemClick="handle_radioButtonGroup_itemClick();"/>
                            <mx:RadioButton
groupName="radioButtonGroupRows" id="radioButtonEmptyRows" selected="true"
y="0"/>
                            <mx:RadioButton
groupName="radioButtonGroupRows" id="radioButtonClustersOnRows" y="20"/>
                            <mx:RadioButton
groupName="radioButtonGroupRows" id="radioButtonExchangesOnRows" y="40"/>
                            <mx:RadioButton
groupName="radioButtonGroupRows" id="radioButtonLiquidityOnRows" y="60"/>
                            <mx:RadioButton
groupName="radioButtonGroupRows" id="radioButtonMarketCapOnRows" y="80"/>
                        </mx:Canvas>
                        <mx:Canvas x="149" y="20">
                            <mx:RadioButtonGroup
id="radioButtonGroupColumns"
itemClick="handle_radioButtonGroup_itemClick();"/>
                            <mx:RadioButton
groupName="radioButtonGroupColumns" id="radioButtonEmptyColumns"
selected="true" y="0"/>
                            <mx:RadioButton
groupName="radioButtonGroupColumns" id="radioButtonClustersOnColumns"
y="20"/>
                            <mx:RadioButton
groupName="radioButtonGroupColumns" id="radioButtonExchangesOnColumns"
y="40"/>
                            <mx:RadioButton
groupName="radioButtonGroupColumns" id="radioButtonLiquidityOnColumns"
y="60"/>
                            <mx:RadioButton
groupName="radioButtonGroupColumns" id="radioButtonMarketCapOnColumns"
y="80"/>
                        </mx:Canvas>
                    </mx:Panel>
                    <mx:Panel height="121" id="panelClusters" layout="absolute"
title="Clusters" width="200">
```

```
                        <controls:Collapser
click="handle_collapser_click(collpserClusters, panelClusters, 127);"
id="collpserClusters" toolTip="Collapse" x="181" y="1"/>
                        <mx:Canvas height="100%"
id="canvasClusterTierSettings" width="100%">
                        <mx:CheckBox
change="handle_checkBoxCluster_change(0);" id="checkBoxCluster0"
label="Cyclicals" selected="true" x="10" y="2"/>
                        <mx:CheckBox
change="handle_checkBoxCluster_change(1);" id="checkBoxCluster1"
label="Energy" selected="true" x="10" y="20"/>
                        <mx:CheckBox
change="handle_checkBoxCluster_change(2);" id="checkBoxCluster2"
label="Financials" selected="true" x="10" y="38"/>
                        <mx:CheckBox
change="handle_checkBoxCluster_change(3);" id="checkBoxCluster3"
label="Healthcare" selected="true" x="10" y="56"/>
                        <mx:CheckBox
change="handle_checkBoxCluster_change(4);" id="checkBoxCluster4"
label="Industrials" selected="true" x="10" y="74"/>
                        <mx:CheckBox
change="handle_checkBoxCluster_change(5);" id="checkBoxCluster5"
label="Materials" selected="true" x="92" y="2"/>
                        <mx:CheckBox
change="handle_checkBoxCluster_change(6);" id="checkBoxCluster6" label="Non-
Cyclicals" selected="true" x="92" y="20"/>
                        <mx:CheckBox
change="handle_checkBoxCluster_change(7);" id="checkBoxCluster7"
label="Technology" selected="true" x="92" y="38"/>
                        <mx:CheckBox
change="handle_checkBoxCluster_change(8);" id="checkBoxCluster8"
label="Telecom" selected="true" x="92" y="56"/>
                        <mx:CheckBox
change="handle_checkBoxCluster_change(9);" id="checkBoxCluster9"
label="Utilities" selected="true" x="92" y="74"/>
                        </mx:Canvas>
                    </mx:Panel>
                    <mx:Panel height="49" id="panelExchanges" layout="absolute"
title="Exchanges" width="200">
                        <controls:Collapser
click="handle_collapser_click(collpserExchanges, panelExchanges, 50);"
id="collpserExchanges" toolTip="Collapse" x="181" y="1"/>
                        <mx:Canvas height="100%"
id="canvasExchangeTierSettings" width="100%">
                        <mx:CheckBox
change="handle_checkBoxExchange_change(EnumExchanges.AMEX);"
id="checkBoxExchangeAMEX" label="AMEX" selected="true" x="10" y="2"/>
                        <mx:CheckBox
change="handle_checkBoxExchange_change(EnumExchanges.NSDQ);"
id="checkBoxExchangeNSDQ" label="NASD" selected="true" x="68" y="2"/>
                        <mx:CheckBox
change="handle_checkBoxExchange_change(EnumExchanges.NYSE);"
id="checkBoxExchangeNYSE" label="NYSE" selected="true" x="128" y="2"/>
                        </mx:Canvas>
                    </mx:Panel>
```

```
                    <mx:Panel height="81" horizontalScrollPolicy="off"
id="panelLiquidity" layout="absolute" title="Liquidity"
verticalScrollPolicy="off" width="200">
                        <controls:Collapser
click="handle_collapser_click(collpserLiquidity, panelLiquidity, 83);"
id="collpserLiquidity" toolTip="Collapse" x="181" y="1"/>
                        <mx:VBox height="62" horizontalScrollPolicy="off"
verticalScrollPolicy="off" width="100%">
                            <mx:Canvas height="60"
horizontalScrollPolicy="off" id="canvasLiquidityFilterSettings"
verticalScrollPolicy="off" width="100%" >
                                <!-- Dual slider will be inserted here --
>
                            </mx:Canvas>
                            <mx:Canvas height="30"
horizontalScrollPolicy="off" id="canvasLiquidityTierSettings"
verticalScrollPolicy="off" visible="false" width="100%">
                                <mx:Label x="10" y="8" text="Show"/>
                                <mx:ComboBox
change="handle_comboBoxLiquidity_change();" id="comboBoxLiquidity"
rowCount="9" selectedIndex="1" width="61" x="51" y="6">
                                    <mx:ArrayCollection>
                                        <mx:String>2</mx:String>
                                        <mx:String>3</mx:String>
                                        <mx:String>4</mx:String>
                                        <mx:String>5</mx:String>
                                        <mx:String>6</mx:String>
                                        <mx:String>7</mx:String>
                                        <mx:String>8</mx:String>
                                        <mx:String>9</mx:String>
                                        <mx:String>10</mx:String>
                                    </mx:ArrayCollection>

                                </mx:ComboBox>
                                <mx:Label text="tiers" x="120" y="8"/>
                            </mx:Canvas>
                        </mx:VBox>
                    </mx:Panel>
                    <mx:Panel height="81" horizontalScrollPolicy="off"
id="panelMarketCap" layout="absolute" title="Market Cap"
verticalScrollPolicy="off" width="200">
                        <controls:Collapser
click="handle_collapser_click(collpserMarketCap, panelMarketCap, 83);"
id="collpserMarketCap" toolTip="Collapse" x="181" y="1"/>
                        <mx:VBox height="62" horizontalScrollPolicy="off"
verticalScrollPolicy="off" width="100%">
                            <mx:Canvas height="60"
horizontalScrollPolicy="off" id="canvasMarketCapFilterSettings"
verticalScrollPolicy="off" width="100%">
                                <!-- Dual slider will be inserted here --
>
                            </mx:Canvas>
                            <mx:Canvas height="30"
horizontalScrollPolicy="off" id="canvasMarketCapTierSettings"
verticalScrollPolicy="off" visible="false" width="100%">
                                <mx:Label x="10" y="8" text="Show"/>
```

```xml
                              <mx:ComboBox
change="handle_comboBoxMarketCap_change();" id="comboBoxMarketCap"
rowCount="9" selectedIndex="1" width="61" x="51" y="6">
                         <mx:ArrayCollection>
                              <mx:String>2</mx:String>
                              <mx:String>3</mx:String>
                              <mx:String>4</mx:String>
                              <mx:String>5</mx:String>
                              <mx:String>6</mx:String>
                              <mx:String>7</mx:String>
                              <mx:String>8</mx:String>
                              <mx:String>9</mx:String>
                              <mx:String>10</mx:String>
                         </mx:ArrayCollection>

                    </mx:ComboBox>
                    <mx:Label text="tiers" x="120" y="8"/>
               </mx:Canvas>
          </mx:VBox>
     </mx:Panel>
     <mx:Panel height="58" horizontalScrollPolicy="off"
id="panelSignal" layout="absolute" status=">0" title="Signal"
verticalScrollPolicy="off" width="200">
               <controls:Collapser id="collpserSignal"
click="handle_collapser_click(collpserSignal, panelSignal, 61);"
toolTip="Collapse" x="181" y="1"/>
               <controls:GlyphSignalScale height="6" width="156"
x="18" y="23"/>
               <mx:HSlider change="handle_hSliderSignal_change();"
height="22" id="hSliderSignal" labelOffset="0" labels="[0,1,2,3,4,5,6]"
maximum="6" minimum="0" snapInterval="0.25" value="0" width="170" x="10"
y="3"/>
          </mx:Panel>
          <mx:Panel height="58" horizontalScrollPolicy="off"
id="panelFlashing" layout="absolute" status="Top 15" title="Flashing"
verticalScrollPolicy="off" width="200">
               <controls:Collapser id="collpserFlashing"
click="handle_collapser_click(collpserFlashing, panelFlashing, 61);"
toolTip="Collapse" x="181" y="1"/>
               <mx:HSlider change="handle_hSliderFlashing_change();"
dataTipPrecision="0" height="22" id="hSliderFlashing" labelOffset="0"
labels="[0,5,10,15,20,25]" maximum="25" minimum="0" snapInterval="1"
value="15" width="170" x="10" y="3"/>
          </mx:Panel>
          <mx:Panel height="135" id="panelTickerInfo"
layout="absolute" title="Ticker" verticalScrollPolicy="off" width="200">
               <controls:Collapser
click="handle_collapser_click(collpserTickerInfo, panelTickerInfo, 187);"
id="collpserTickerInfo" toolTip="Collapse" x="181" y="1"/>
               <mx:Form width="170" x="10" y="2">
                    <mx:FormItem id="formItemSymbol"
label="Symbol:" width="100%">
                              <mx:Label id="formItemSymbolLabel"
text=""/>
                    </mx:FormItem>
                    <mx:FormItem id="formItemSignal"
label="Signal:" width="100%">
```

```xml
                                    <mx:Label id="formItemSignalLabel"
text=""/>
                                </mx:FormItem>
                                <mx:FormItem id="formItemExchange"
label="Exchange:" width="100%">
                                    <mx:Label id="formItemExchangeLabel"
text=""/>
                                </mx:FormItem>
                                <mx:FormItem id="formItemCluster"
label="Cluster:" width="100%">
                                    <mx:Label id="formItemClusterLabel"
text=""/>
                                </mx:FormItem>
                                <mx:FormItem id="formItemLiquidity"
label="Liquidity:" width="100%">
                                    <mx:Label id="formItemLiquidityLabel"
text=""/>
                                </mx:FormItem>
                                <mx:FormItem id="formItemMarketCap"
label="Market Cap:" width="100%">
                                    <mx:Label id="formItemMarketCapLabel"
text=""/>
                                </mx:FormItem>
                            </mx:Form>
                            <controls:GlyphSignalIndicator height="11"
id="glyphSignalIndicator" visible="false" width="20" x="139" y="23"/>
                        </mx:Panel>
                    </mx:VBox>
                    <mx:Canvas horizontalScrollPolicy="off" height="30"
id="canvasFind" styleName="canvasFind" verticalScrollPolicy="off" width="200"
x="20">
                        <mx:HBox x="14" y="6">
                            <mx:TextInput height="16" id="textInputFind"
keyDown="handle_textInputFind_keyDown(event);" maxChars="6" restrict="A-Z."
width="63"/>
                            <mx:Button click="handle_buttonFind_click();"
height="16" id="buttonFind" label="Find"/>
                        </mx:HBox>
                        <controls:CloseButton click="currentState=null;"
toolTip="Collapse" x="185" y="28"/>
                        <mx:Label id="labelInvalidTicker" visible="false" x="16"
y="28"/>
                        <mx:Form id="formFind" visible="false" width="170" x="14"
y="28">
                            <mx:FormItem id="formItemSymbolPopup" label="Symbol:"
width="100%">
                                <mx:Label id="formItemSymbolLabelPopup"
text=""/>
                            </mx:FormItem>
                            <mx:FormItem id="formItemSignalPopup" label="Signal:"
width="100%">
                                <mx:Label id="formItemSignalLabelPopup"
text=""/>
                            </mx:FormItem>
                            <mx:FormItem id="formItemExchangePopup"
label="Exchange:" width="100%">
```

```xml
                                <mx:Label id="formItemExchangeLabelPopup"
text=""/>
                            </mx:FormItem>
                            <mx:FormItem id="formItemClusterPopup"
label="Cluster:" width="100%">
                                <mx:Label id="formItemClusterLabelPopup"
text=""/>
                            </mx:FormItem>
                            <mx:FormItem id="formItemLiquidityPopup"
label="Liquidity:" width="100%">
                                <mx:Label id="formItemLiquidityLabelPopup"
text=""/>
                            </mx:FormItem>
                            <mx:FormItem id="formItemMarketCapPopup"
label="Market Cap:" width="100%">
                                <mx:Label id="formItemMarketCapLabelPopup"
text=""/>
                            </mx:FormItem>
                        </mx:Form>
                        <controls:GlyphSignalIndicator height="11"
id="glyphSignalIndicatorPopup" visible="false" width="20" x="139" y="49"/>
                </mx:Canvas>
                <controls:Matrix height="100%" id="matrix" left="241" top="0"
visible="false" width="100%"/>
            </mx:Canvas>
</mx:Application>
// END FILE "/VMap/src/VMap.mxml"

// BEGIN FILE "/VMap/src/com/vynance/app/AppManager.as"
package com.vynance.app
{
        import com.vynance.model.Map;
        import com.vynance.model.ParamCluster;
        import com.vynance.model.ParamExchange;
        import com.vynance.model.ParamLiquidity;
        import com.vynance.model.ParamMarketCap;
        import com.vynance.model.ParamSignal;

        import flash.events.EventDispatcher;
        import flash.system.Security;

        public class AppManager extends EventDispatcher
        {
                public static const DEBUG:Boolean = false;

                private static var _instance:AppManager = new AppManager();

                public static function getInstance():AppManager
                {
                        return _instance;
                }

                public var flashingRank:int;

                public var log:String;
```

```actionscript
public var mainCanvasResized:Boolean;

public var mapDownloaded:Boolean;

public var marketStatus:int;

public var matrixColumns:int;

public var matrixRows:int;

public var signalGraphWidth:int = 195;

private var _map:Map;

private var _paramCluster:ParamCluster;

private var _paramExchange:ParamExchange;

private var _paramLiquidity:ParamLiquidity;

private var _paramMarketCap:ParamMarketCap;

private var _paramSignal:ParamSignal;

public function AppManager()
{
    if (_instance != null)
        throw("AppManager is already created");

    flashingRank = 15;

    _map = new Map();

    _paramCluster = new ParamCluster;
    _paramExchange = new ParamExchange;
    _paramLiquidity = new ParamLiquidity;
    _paramMarketCap = new ParamMarketCap;
    _paramSignal = new ParamSignal;

    Security.allowDomain("www.{domain}.com/{directory}");
    Security.allowDomain("www.{domain}.com/{directory}/m.txt");
// COMMENT: HERE {domain} AND {directory} ARE PLACEHOLDERS TO BE MODIFIED
}

public function get map():Map
{
    return _map;
}

public function get paramCluster():ParamCluster
{
    return _paramCluster;
}

public function get paramExchange():ParamExchange
{
    return _paramExchange;
```

```
                }

                public function get paramLiquidity():ParamLiquidity
                {
                        return _paramLiquidity;
                }

                public function get paramMarketCap():ParamMarketCap
                {
                        return _paramMarketCap;
                }

                public function get paramSignal():ParamSignal
                {
                        return _paramSignal;
                }
        }
}
// END FILE "/VMap/src/com/vynance/app/AppManager.as"

// BEGIN FILE "/VMap/src/com/vynance/app/URLDownloader.as"
package com.vynance.app
{
    import flash.events.*;
    import flash.net.*;

    public class URLDownloader extends EventDispatcher
    {
            private var _urlDownloaderEvent:URLDownloaderEvent;

            private var _urlLoader:URLLoader;

            private var _urlRequest:URLRequest;

            public function URLDownloader()
            {
                    _urlLoader = new URLLoader();
                    _urlRequest = new URLRequest();

        IEventDispatcher(_urlLoader).addEventListener(Event.COMPLETE,
handle_COMPLETE);
            }

            public function get urlLoader():URLLoader
            {
                    return _urlLoader;
            }

            public function download(url:String):void
            {
                    _urlRequest.url = url + "?r=" + int(Math.random() * 1000);

        IEventDispatcher(_urlLoader).addEventListener(Event.COMPLETE,
handle_COMPLETE);
```

```actionscript
        IEventDispatcher(_urlLoader).addEventListener(HTTPStatusEvent.HTTP_STAT
US, handle_HTTP_STATUS);

        IEventDispatcher(_urlLoader).addEventListener(IOErrorEvent.IO_ERROR,
handle_IO_ERROR);
                IEventDispatcher(_urlLoader).addEventListener(Event.OPEN,
handle_OPEN);

        IEventDispatcher(_urlLoader).addEventListener(ProgressEvent.PROGRESS,
handle_PROGRESS);

        IEventDispatcher(_urlLoader).addEventListener(SecurityErrorEvent.SECURI
TY_ERROR, handle_SECURITY_ERROR);

                try
                {
                        _urlLoader.load(_urlRequest);
                }
                catch (error:Error)
                {
                }
        }

        private function handle_COMPLETE(event:Event):void
        {
                dispatchEvent(new
URLDownloaderEvent(URLDownloaderEvent.COMPLETE));
        }

        private function handle_HTTP_STATUS(event:HTTPStatusEvent):void
        {
        }

        private function handle_IO_ERROR(event:IOErrorEvent):void
        {
        }

        private function handle_OPEN(event:Event):void
        {
        }

        private function handle_PROGRESS(event:ProgressEvent):void
        {
        }

        private function
handle_SECURITY_ERROR(event:SecurityErrorEvent):void
        {
        }
    }
}
// END FILE "/VMap/src/com/vynance/app/URLDownloader.as"

// BEGIN FILE "/VMap/src/com/vynance/app/URLDownloaderEvent.as"
package com.vynance.app
{
```

```actionscript
        import flash.events.Event;

        public class URLDownloaderEvent extends Event
        {
                public static const COMPLETE:String = "COMPLETE";

                public function URLDownloaderEvent(type:String)
                {
                        super(type, true);
                }
        }
}
// END FILE "/VMap/src/com/vynance/app/URLDownloaderEvent.as"

// BEGIN FILE "/VMap/src/com/vynance/controls/bubble/BubbleContainer.as"
package com.vynance.controls.bubble
{
        import com.vynance.app.AppManager;
        import com.vynance.controls.GlyphTicker;
        import com.vynance.model.Map;
        import com.vynance.model.MapEvent;
        import com.vynance.model.Ticker;
        import com.vynance.modules.EnumMarketStatuses;

        import flash.events.Event;

        import mx.controls.Label;
        import mx.core.UIComponent;

        public class BubbleContainer extends UIComponent
        {
                private const GLYPH_COUNT:int = 80;

                private const MARGIN_BOTTOM:int = 25;

                private const MARGIN_LEFT:int = 25;

                private const MARGIN_RIGHT:int = 25;

                private const MARGIN_TOP:int = 25;

                private const MATRIX_MIN_HEIGHT:int = 290;

                private const MATRIX_MIN_WIDTH:int = 621;

                private var _bubbleGlyphs:Array;

                private var _height:int;

                private var _labelSignal10:Label;

                private var _labelSignal50:Label;

                private var _labelSignalNegative10:Label;

                private var _labelSignalNegative50:Label;
```

```actionscript
        private var _labelUpdating:Label;

        private var _width:int;

        public function BubbleContainer()
        {
                super();

    AppManager.getInstance().map.addEventListener(MapEvent.BEGIN_SIGNAL_DOW
NLOAD, handle_Map_BeginSignalDownload);

    AppManager.getInstance().map.addEventListener(MapEvent.END_SIGNAL_DOWNL
OAD, handle_Map_EndSignalDownload);
        }

        override protected function commitProperties():void
        {
                if (this.height == 0)
                        return;

                var currentGlyphIndex:int;
                var glyph:GlyphBubble;
                var i:int;
                var map:Map = AppManager.getInstance().map;
                var ticker:Ticker;
                var signalIndex:int;

                if (map.tickerCount() == 0)
                        return;

                _height = this.height;
                if (_height < MATRIX_MIN_HEIGHT)
                        _height = MATRIX_MIN_HEIGHT;
                _width = this.width;
                if (_width < MATRIX_MIN_WIDTH)
                        _width = MATRIX_MIN_WIDTH;

                signalIndex = map.tickerCount() - 1;
                while (signalIndex >= 0 && currentGlyphIndex < GLYPH_COUNT)
{
                        ticker =
map.tickerByIndex(map.tickerIndexBySignalIndex(signalIndex));
                        if (!ticker.excluded) {
                                if (!isNaN(ticker.signalChange) ||
ticker.signalChange == 0) {
                                        glyph =
GlyphBubble(_bubbleGlyphs[currentGlyphIndex]);
                                        glyph.tickerIndex =
map.tickerIndexBySignalIndex(signalIndex);
                                        glyph.visible = true;
                                        currentGlyphIndex++;
                                }
                        }
                        signalIndex--;
                }
```

```actionscript
            for (i = currentGlyphIndex; i < GLYPH_COUNT; i++) {
                GlyphBubble(_bubbleGlyphs[i]).visible = false;
            }
    }

    override protected function createChildren():void
    {
        var glyph:GlyphBubble;

        _labelUpdating = new Label();
        _labelUpdating.height = 17;
        _labelUpdating.selectable = false;
        _labelUpdating.setStyle("color", "#888888");
        _labelUpdating.text= "Updating...";
        _labelUpdating.width = 100;
        _labelUpdating.visible = false;
        this.addChild(_labelUpdating);

        _labelSignal10 = new Label();
        _labelSignal10.height = 17;
        _labelSignal10.selectable = false;
        _labelSignal10.setStyle("color", "#888888");
        _labelSignal10.setStyle("textAlign", "center");
        _labelSignal10.text = "10";
        _labelSignal10.width = 50;
        this.addChild(_labelSignal10);

        _labelSignal50 = new Label();
        _labelSignal50.height = 17;
        _labelSignal50.selectable = false;
        _labelSignal50.setStyle("color", "#888888");
        _labelSignal50.setStyle("textAlign", "center");
        _labelSignal50.text = "50";
        _labelSignal50.width = 50;
        this.addChild(_labelSignal50);

        _labelSignalNegative10 = new Label();
        _labelSignalNegative10.height = 17;
        _labelSignalNegative10.selectable = false;
        _labelSignalNegative10.setStyle("color", "#888888");
        _labelSignalNegative10.setStyle("textAlign", "center");
        _labelSignalNegative10.text = "-10";
        _labelSignalNegative10.width = 50;
        this.addChild(_labelSignalNegative10);

        _labelSignalNegative50 = new Label();
        _labelSignalNegative50.height = 17;
        _labelSignalNegative50.selectable = false;
        _labelSignalNegative50.setStyle("color", "#888888");
        _labelSignalNegative50.setStyle("textAlign", "center");
        _labelSignalNegative50.text = "-50";
        _labelSignalNegative50.width = 50;
        this.addChild(_labelSignalNegative50);

        _bubbleGlyphs = new Array();
        for (var i:int = 0; i < GLYPH_COUNT; i++) {
            glyph = new GlyphBubble();
```

```actionscript
                        this.addChild(glyph);
                        _bubbleGlyphs.push(glyph);
                    }
                }
            }

            override protected function
updateDisplayList(unscaledWidth:Number, unscaledHeight:Number):void
            {
                    var coordinate:int;
                    var glyph:GlyphBubble;
                    var map:Map = AppManager.getInstance().map
                    var ticker:Ticker;

                    this.graphics.clear();

                    this.graphics.lineStyle(1, 0x888888);
                    this.graphics.moveTo(unscaledWidth / 2, MARGIN_TOP);
                    this.graphics.lineTo(unscaledWidth / 2, unscaledHeight -
MARGIN_BOTTOM);
                    this.graphics.moveTo(MARGIN_LEFT, unscaledHeight / 2);
                    this.graphics.lineTo(unscaledWidth - MARGIN_RIGHT,
unscaledHeight / 2);

                    coordinate = calculateX(unscaledWidth, -50);
                    this.graphics.moveTo(coordinate, unscaledHeight / 2 - 5);
                    this.graphics.lineTo(coordinate, unscaledHeight / 2 + 5);
                    _labelSignalNegative50.x = coordinate -
_labelSignalNegative50.width / 2;
                    _labelSignalNegative50.y = unscaledHeight / 2 + 5;

                    coordinate = calculateX(unscaledWidth, -10);
                    this.graphics.moveTo(coordinate, unscaledHeight / 2 - 5);
                    this.graphics.lineTo(coordinate, unscaledHeight / 2 + 5);
                    _labelSignalNegative10.x = coordinate -
_labelSignalNegative10.width / 2;
                    _labelSignalNegative10.y = unscaledHeight / 2 + 5;

                    coordinate = calculateX(unscaledWidth, 10);
                    this.graphics.moveTo(coordinate, unscaledHeight / 2 - 5);
                    this.graphics.lineTo(coordinate, unscaledHeight / 2 + 5);
                    _labelSignal10.x = coordinate - _labelSignal10.width / 2;
                    _labelSignal10.y = unscaledHeight / 2 + 5;

                    coordinate = calculateX(unscaledWidth, 50);
                    this.graphics.moveTo(coordinate, unscaledHeight / 2 - 5);
                    this.graphics.lineTo(coordinate, unscaledHeight / 2 + 5);
                    _labelSignal50.x = coordinate - _labelSignal50.width / 2;
                    _labelSignal50.y = unscaledHeight / 2 + 5;

                    if (map.tickerCount() == 0)
                            return;

                    for (var i:int = 0; i < GLYPH_COUNT; i++) {
                            glyph = GlyphBubble(_bubbleGlyphs[i]);
                            if (glyph.visible) {
                                    ticker = map.tickerByIndex(glyph.tickerIndex);
                                    glyph.height = GlyphTicker.GLYPH_TICKER_WIDTH;
```

```actionscript
                            glyph.width = GlyphTicker.GLYPH_TICKER_WIDTH;
                            glyph.x = calculateX(unscaledWidth,
ticker.signal);
                            glyph.y = calculateY(unscaledHeight,
ticker.signalChange);
                            glyph.invalidateDisplayList();
                    }
                }
            }

            private function handle_Map_BeginSignalDownload(event:Event):void
            {
                    if (AppManager.getInstance().marketStatus ==
EnumMarketStatuses.OPEN)
                            _labelUpdating.visible = true;
            }

            private function handle_Map_EndSignalDownload(event:Event):void
            {
                    _labelUpdating.visible = false;

                    callLater(this.invalidateProperties);
                    callLater(this.invalidateDisplayList);
            }

            private function calculateX(totalWidth:int, signal:Number):int
            {
                    var highRangeWidth:int = (totalWidth - MARGIN_LEFT -
MARGIN_RIGHT) / 12;

                    if (signal < -50)
                            return MARGIN_LEFT;

                    if (signal < -10)
                            return MARGIN_LEFT + highRangeWidth + highRangeWidth
* (signal + 10) / (50 - 10);

                    if (signal > 50)
                            return totalWidth - MARGIN_RIGHT;

                    if (signal > 10)
                            return totalWidth - MARGIN_RIGHT - highRangeWidth +
highRangeWidth * (signal - 10) / (50 - 10);

                    return totalWidth / 2 + (signal / 10) * (totalWidth / 2 -
MARGIN_LEFT - highRangeWidth);
            }

            private function calculateY(totalHeight:int,
signalChange:Number):int
            {
                    const MAX_VALUE:Number = .5;

                    if (signalChange < -MAX_VALUE)
                            return MARGIN_TOP;

                    if (signalChange > MAX_VALUE)
```

```
                        return totalHeight - MARGIN_BOTTOM;

                    return totalHeight / 2 + (signalChange / MAX_VALUE) *
(totalHeight / 2 - MARGIN_TOP);
                }
            }
        }
    }
}
// END FILE "/VMap/src/com/vynance/controls/bubble/BubbleContainer.as"

// BEGIN FILE "/VMap/src/com/vynance/controls/bubble/GlyphBubble.as"
package com.vynance.controls.bubble
{
        import com.vynance.app.AppManager;
        import com.vynance.controls.GlyphTickerEvent;
        import com.vynance.model.Ticker;
        import com.vynance.utils.Signal;

        import flash.events.MouseEvent;
        import flash.text.TextField;
        import flash.text.TextFormat;

        import mx.core.UIComponent;

        public class GlyphBubble extends UIComponent
        {
                public static var format:TextFormat;

                public static var formatHovered:TextFormat;

                public static const GLYPH_TICKER_HEIGHT:int = 17;

                public static const GLYPH_TICKER_WIDTH:int = 43;

                private var _hovered:Boolean;

                private var _label:TextField;

                private var _tickerIndex:int;

                public function GlyphBubble()
                {
                        super();

                        _tickerIndex = -1;

                        if (format == null) {
                                format = new TextFormat();
                                format.align = "center";
                                format.bold = true;
                                format.color = 0xcccccc;
                                format.font = "Arial";
                                format.size = 10;
                        }

                        if (formatHovered == null) {
                                formatHovered = new TextFormat();
                                formatHovered.align = "center";
```

```actionscript
                        formatHovered.bold = true;
                        formatHovered.color = 0xffffff;
                        formatHovered.font = "Arial";
                        formatHovered.size = 10;
                    }

                    _label = new TextField;
                    _label.defaultTextFormat = format;
                    _label.height = 17;
                    _label.selectable = false;
                    _label.visible = true;
                    _label.width = GlyphBubble.GLYPH_TICKER_WIDTH;
                    _label.x = 0;
                    _label.y = 0;
                    this.addChild(_label);

                    this.addEventListener(MouseEvent.ROLL_OUT,
handle_ROLL_OUT);
                    this.addEventListener(MouseEvent.ROLL_OVER,
handle_ROLL_OVER);
                }

            public function get tickerIndex():int
            {
                    return _tickerIndex;
            }

            public function set tickerIndex(value:int):void
            {
                    _tickerIndex = value;
            }

            override protected function
updateDisplayList(unscaledWidth:Number, unscaledHeight:Number):void
            {
                    var color:int;
                    var radius:int;
                    var thickness:int;
                    var ticker:Ticker;

                    if (_tickerIndex == -1) return;

                    ticker =
AppManager.getInstance().map.tickerByIndex(_tickerIndex);

                    this.graphics.clear();

                    if (_hovered)
                        _label.defaultTextFormat = GlyphBubble.formatHovered;
                    else
                        _label.defaultTextFormat = GlyphBubble.format;
                    _label.text = ticker.symbol;
                    _label.width = this.unscaledWidth;
                    _label.x = -this.unscaledWidth / 2;
                    _label.y = -this.unscaledHeight / 2 + 12;

                    radius = 15 + ticker.liquidity.toString().length * 2;
```

```
                    if (ticker.marketCap.toString().length < 7)
                        thickness = 1;
                    else
                        thickness = ticker.marketCap.toString().length - 5;

                    this.graphics.beginFill(Signal.getColor(ticker));
                    this.graphics.drawCircle(0, 0, radius);
                    this.graphics.endFill();

                    if (_hovered) {
                        this.graphics.beginFill(0xffffff);
                        this.graphics.drawCircle(0, 0, 15);
                        this.graphics.endFill();
                    }

                    this.alpha = 0.20;
                }

            private function handle_ROLL_OUT(event:MouseEvent):void
            {
                _hovered = false;
                this.invalidateDisplayList();
                this.dispatchEvent(new
GlyphTickerEvent(GlyphTickerEvent.ROLL_OUT, _tickerIndex));
            }

            private function handle_ROLL_OVER(event:MouseEvent):void
            {
                _hovered = true;
                this.invalidateDisplayList();
                this.dispatchEvent(new
GlyphTickerEvent(GlyphTickerEvent.ROLL_OVER, _tickerIndex));
            }
        }
}
// END FILE "/VMap/src/com/vynance/controls/bubble/GlyphBubble.as"

// BEGIN FILE "/VMap/src/com/vynance/controls/dualSlider/DualSlider.as"
package com.vynance.controls.dualSlider
{
    import flash.display.DisplayObject;
    import flash.events.Event;
    import flash.events.MouseEvent;
    import flash.geom.Rectangle;

    import mx.controls.Label;
    import mx.core.UIComponent;

    public class DualSlider extends UIComponent
    {
        private const THUMB_HEIGHT:int = 13;

        private const THUMB_WIDTH:int = 9;

        private const THUMB_Y:int = 29;
```

```actionscript
            private const TRACK_TOP:int = 21;

            private const H_MARGIN:int = 10;

            private var _captions:Array;

            private var _height:int;

            private var _labels:Array;

            private var _left:Thumb;

            private var _maxIndex:int;

            private var _maxLabel:Label;

            private var _minIndex:int;

            private var _minLabel:Label;

            private var _right:Thumb;

            private var _trackHighlight:TrackHighlight;

            private var _values:Array;

            private var _width:int;

            public function DualSlider(values:Array, captions:Array,
markers:Array)
            {
                _values = new Array();
                for (var i:int = 0; i < values.length; i++) {
                    _values[i] = values[i];
                }

                _captions = new Array();
                for (i = 0; i < captions.length; i++) {
                    _captions[i] = captions[i];
                }

                _labels = new Array;
                for (i = 0; i < markers.length; i++) {
                    if (markers[i] != null) {
                        var label:Label = new Label();
                        label.height = 17;
                        label.setStyle("fontFamily", "Arial");
                        label.setStyle("fontSize", 9);
                        label.text = markers[i];
                        this.addChild(label);
                        _labels[i] = label;
                    }
                }

                _minIndex = 0;
                _maxIndex = _values.length - 1;
```

```actionscript
                _left = new Thumb(true);
                this.addChild(_left as DisplayObject);

                _right = new Thumb(false);
                this.addChild(_right as DisplayObject);

                _trackHighlight = new TrackHighlight();
                this.addChild(_trackHighlight);

                _minLabel = new Label();
                _minLabel.setStyle("textAlign", "left");
                _minLabel.text = _captions[_minIndex];
                this.addChild(_minLabel);

                _maxLabel = new Label();
                _maxLabel.setStyle("textAlign", "right");
                _maxLabel.text = _captions[_maxIndex];
                this.addChild(_maxLabel);
        }

        override protected function commitProperties():void
        {
                trace("DualSlider::commitProperties " + _width + " " +
_height);
                _height = this.height;
                _width = this.width;

                _left.height = THUMB_HEIGHT;
                _left.width = THUMB_WIDTH;
                _left.x = getXFromValue(_minIndex) - (THUMB_WIDTH - 1);
                _left.y = THUMB_Y;

                _right.height = THUMB_HEIGHT;
                _right.width = THUMB_WIDTH;
                _right.x = getXFromValue(_maxIndex);
                _right.y = THUMB_Y;
        }

        override protected function measure():void
        {
                trace("DualSlider::measure " + _width + " " + _height);
        }

        override protected function
updateDisplayList(unscaledWidth:Number, unscaledHeight:Number):void
        {
                trace("DualSlider::updateDisplayList " + _width + " " +
_height);
                var label:Label;
                var xCoor:int;

                if (_height == 0) return;
                if (_width == 0) return;

                graphics.clear();

                graphics.beginFill(0x404040);
```

```
                    graphics.drawRect(0, 0, _width, _height);
                    graphics.endFill();

                    for (var i:int = 0; i < _values.length; i++) {
                        xCoor = getXFromValue(i);
                        if (_labels[i] != null) {
                            label = Label(_labels[i]);
                            label.width = label.textWidth + 6;
                            label.x = getXFromValue(i) - (label.textWidth +
6)  / 2;

                            label.y = TRACK_TOP - 15;
                            graphics.lineStyle(0);
                            graphics.moveTo(xCoor, TRACK_TOP - 3);
                            graphics.lineStyle(1, 0x888888);
                            graphics.lineTo(xCoor, TRACK_TOP + 6);
                        }
                    }

                    graphics.moveTo(H_MARGIN, TRACK_TOP);
                    graphics.lineStyle(1, 0xcccccc);
                    graphics.lineTo(_width - H_MARGIN, TRACK_TOP);

                    graphics.moveTo(H_MARGIN, TRACK_TOP + 1);
                    graphics.lineStyle(1, 0xeeeeee);
                    graphics.lineTo(_width - H_MARGIN, TRACK_TOP + 1);

                    graphics.moveTo(H_MARGIN - 1, TRACK_TOP + 1);
                    graphics.lineStyle(1, 0xcccccc);
                    graphics.lineTo(H_MARGIN, TRACK_TOP + 2);
                    graphics.lineTo(_width - H_MARGIN, TRACK_TOP + 2);
                    graphics.lineTo(_width - H_MARGIN + 1, TRACK_TOP + 1);

                    graphics.moveTo(H_MARGIN - 1, TRACK_TOP + 2);
                    graphics.lineStyle(1, 0x888888);
                    graphics.lineTo(H_MARGIN, TRACK_TOP + 3);
                    graphics.lineTo(_width - H_MARGIN, TRACK_TOP + 3);
                    graphics.lineTo(_width - H_MARGIN + 1, TRACK_TOP + 2);

                    _minLabel.height = 17;
                    _minLabel.width = _width / 2 - H_MARGIN;
                    _minLabel.x = H_MARGIN - 3;
                    _minLabel.y = 40;

                    _maxLabel.height = 17;
                    _maxLabel.width = _width / 2 - H_MARGIN + 3;
                    _maxLabel.x = _width / 2;
                    _maxLabel.y = 40;

                    moveTrackHighlight();
                }

            public function onThumbMouseDown(thumb:Thumb):void
            {
                    var xMin:int;
                    var xMax:int;

                    if (thumb == _left) {
```

```actionscript
                    xMin = getXFromValue(0);
                    xMax = getXFromValue(_maxIndex - 1);
                    _left.startDrag(false, new Rectangle(xMin -
(THUMB_WIDTH - 1), THUMB_Y, xMax - xMin, 0));
                }
                else {
                    xMin = getXFromValue(_minIndex + 1);
                    xMax = getXFromValue(_values.length - 1);
                    _right.startDrag(false, new Rectangle(xMin, THUMB_Y,
xMax - xMin, 0));
                }

                stage.addEventListener(MouseEvent.MOUSE_MOVE,
handle_MOUSE_MOVE);
            }

            public function onThumbMouseUp(thumb:Thumb):void
            {
                thumb.stopDrag();

                if (thumb == _left) {
                    _minIndex = getValueFromX(_left.x + (THUMB_WIDTH -
1));
                    _left.x = getXFromValue(_minIndex) - (THUMB_WIDTH -
1);
                    _minLabel.setStyle("color", "0xff9000");
                }
                else {
                    _maxIndex = getValueFromX(_right.x);
                    _right.x = getXFromValue(_maxIndex);
                    _maxLabel.setStyle("color", "0xff9000");
                }

                moveTrackHighlight();

                stage.removeEventListener(MouseEvent.MOUSE_MOVE,
handle_MOUSE_MOVE);

                this.dispatchEvent(new
DualSliderEvent(DualSliderEvent.CHANGE, _values[_minIndex],
_values[_maxIndex]));
            }

            private function handle_MOUSE_MOVE(event:MouseEvent):void
            {
                if (_left.dragged) {
                    _minIndex = getValueFromX(_left.x + (THUMB_WIDTH -
1))
                    _minLabel.setStyle("color", "0xffffff");
                    _minLabel.text = _captions[_minIndex];
                }
                else if (_right.dragged) {
                    _maxIndex = getValueFromX(_right.x)
                    _maxLabel.setStyle("color", "0xffffff");
                    _maxLabel.text = _captions[_maxIndex];
                }
```

```actionscript
                moveTrackHighlight();
        }

        private function getXFromValue(value:int):int
        {
                return H_MARGIN + (_width - 2 * H_MARGIN - 1) * value /
(_values.length - 1);
        }

        private function getValueFromX(xCoor:int):int
        {
                var value:Number = (xCoor - H_MARGIN) * (_values.length -
1) / (_width - 2 * H_MARGIN - 1);
                return int(Math.round(value));
        }

        private function moveTrackHighlight():void
        {
                var xCoor:int = getXFromValue(_minIndex);

                _trackHighlight.x = xCoor;
                _trackHighlight.y = TRACK_TOP;
                _trackHighlight.draw(getXFromValue(_maxIndex) - xCoor + 1,
4);
        }
    }
}
// END FILE "/VMap/src/com/vynance/controls/dualSlider/DualSlider.as"

// BEGIN FILE "/VMap/src/com/vynance/controls/dualSlider/DualSliderEvent.as"
package com.vynance.controls.dualSlider
{
    import flash.events.Event;

    public class DualSliderEvent extends Event
    {
        public static const CHANGE:String = "CHANGE";

        private var _max:Number;

        private var _min:Number;

        public function DualSliderEvent(type:String, min:Number,
max:Number)
        {
                super(type, true);

                _min = min;
                _max = max;
        }

        public function get max():Number
        {
                return _max;
        }

        public function get min():Number
```

```
                    {
                            return _min;
                    }
            }
}
// END FILE "/VMap/src/com/vynance/controls/dualSlider/DualSliderEvent.as"

// BEGIN FILE "/VMap/src/com/vynance/controls/dualSlider/Thumb.as"
package com.vynance.controls.dualSlider
{
        import flash.events.Event;
        import flash.events.MouseEvent;

        import mx.controls.Image;
        import mx.core.BitmapAsset;
        import mx.core.UIComponent;

        public class Thumb extends UIComponent
        {
                private var _dragged:Boolean;

                private var _height:int;

                private var _image:Image;

               [Embed(source="../assets/leftThumbDown.png")]
                private var imageLeftThumbDown:Class;

                [Embed(source="../assets/leftThumbOver.png")]
                private var imageLeftThumbOver:Class;

                [Embed(source="../assets/leftThumbUp.png")]
                private var imageLeftThumbUp:Class;

                [Embed(source="../assets/rightThumbDown.png")]
                private var imageRightThumbDown:Class;

                [Embed(source="../assets/rightThumbOver.png")]
                private var imageRightThumbOver:Class;

                [Embed(source="../assets/rightThumbUp.png")]
                private var imageRightThumbUp:Class;

                  private var _isLeftThumb:Boolean;

                  private var _width:int;

                  public function Thumb(isLeftThumb:Boolean)
                  {
                          var bitmapAsset:BitmapAsset;

                          _isLeftThumb = isLeftThumb;

                          if (_isLeftThumb)
                                  bitmapAsset = new imageLeftThumbUp() as BitmapAsset;
                          else
                                  bitmapAsset = new imageRightThumbUp() as BitmapAsset;
```

```actionscript
            _image = new Image();
            _image.source = bitmapAsset;
            this.addChild(_image);

            this.addEventListener(MouseEvent.MOUSE_DOWN,
handle_MOUSE_DOWN);
            this.addEventListener(MouseEvent.MOUSE_OUT,
handle_MOUSE_OUT);
            this.addEventListener(MouseEvent.MOUSE_OVER,
handle_MOUSE_OVER);
            this.addEventListener(MouseEvent.MOUSE_UP,
handle_MOUSE_UP);
        }

        public function get dragged():Boolean
        {
            return _dragged;
        }

        override protected function commitProperties():void
        {
            _height = this.height;
            _width = this.width;

            _image.height = _height;
            _image.width = _width;
        }

        private function handle_MOUSE_DOWN(event:MouseEvent):void
        {
            _dragged = true;
            if (_isLeftThumb)
                _image.source = new imageLeftThumbDown() as
BitmapAsset;
            else
                _image.source = new imageRightThumbDown() as
BitmapAsset;
            stage.addEventListener(MouseEvent.MOUSE_UP,
handle_MOUSE_UP);
            DualSlider(owner).onThumbMouseDown(this);
        }

        private function handle_MOUSE_OUT(event:MouseEvent):void
        {
            if (!_dragged) {
                if (_isLeftThumb)
                    _image.source = new imageLeftThumbUp() as
BitmapAsset;
                else
                    _image.source = new imageRightThumbUp() as
BitmapAsset;
            }
        }

        private function handle_MOUSE_OVER(event:MouseEvent):void
        {
```

```actionscript
                    if (!_dragged) {
                        if (_isLeftThumb)
                            _image.source = new imageLeftThumbOver() as
BitmapAsset;
                        else
                            _image.source = new imageRightThumbOver() as
BitmapAsset;
                    }
                }

            private function handle_MOUSE_UP(event:MouseEvent):void
                {
                    _dragged = false;
                    if (_isLeftThumb)
                        _image.source = new imageLeftThumbUp() as
BitmapAsset;
                    else
                        _image.source = new imageRightThumbUp() as
BitmapAsset;
                    stage.removeEventListener(MouseEvent.MOUSE_UP,
handle_MOUSE_UP);
                    DualSlider(owner).onThumbMouseUp(this);
                }
        }
}
// END FILE "/VMap/src/com/vynance/controls/dualSlider/Thumb.as"

// BEGIN FILE "/VMap/src/com/vynance/controls/dualSlider/TrackHighlight.as"
package com.vynance.controls.dualSlider
{
    import flash.display.Sprite;

    public class TrackHighlight extends Sprite
        {
            private const TRACK_COLOR:int = 0x2eaeff;

            public function draw(width:int, height:int):void
                {
                    this.graphics.clear();
                    this.graphics.beginFill(TRACK_COLOR, .6);
                    this.graphics.drawRect(0, 0, width, height);
                }
        }
}
// END FILE "/VMap/src/com/vynance/controls/dualSlider/TrackHighlight.as"

// BEGIN FILE "/VMap/src/com/vynance/controls/CloseButton.as"
package com.vynance.controls
{
    import flash.events.MouseEvent;

    import mx.controls.Image;
    import mx.core.BitmapAsset;
    import mx.core.UIComponent;

    public class CloseButton extends UIComponent
        {
```

```actionscript
        private var _image:Image;

      [Embed(source="../assets/close.png")]
      private var imageClose:Class;

        public function CloseButton()
        {
                var bitmapAsset:BitmapAsset;

                bitmapAsset = new imageClose() as BitmapAsset;
                _image = new Image();
                _image.source = bitmapAsset;
                this.addChild(_image);

                this.buttonMode = true;
        }

        override protected function commitProperties():void
        {
                _image.height = 4;
                _image.width = 8;
        }
    }
}
// END FILE "/VMap/src/com/vynance/controls/CloseButton.as"

// BEGIN FILE "/VMap/src/com/vynance/controls/Collapser.as"
package com.vynance.controls
{
      import mx.controls.Image;
      import mx.core.BitmapAsset;
      import mx.core.UIComponent;

      public class Collapser extends UIComponent
      {
              private var _collapsed:Boolean;

              private var _image:Image;

            [Embed(source="../assets/collapser.png")]
            private var imageCollapser:Class;

            [Embed(source="../assets/expander.png")]
            private var imageExpander:Class;

              public function Collapser()
              {
                      var bitmapAsset:BitmapAsset;

                      bitmapAsset = new imageCollapser() as BitmapAsset;
                      _image = new Image();
                      _image.source = bitmapAsset;
                      this.addChild(_image);

                      this.buttonMode = true;
              }
```

```actionscript
            public function get collapsed():Boolean
            {
                    return _collapsed;
            }

            override protected function commitProperties():void
            {
                    _image.height = 4;
                    _image.width = 8;
            }

            public function toggle():void
            {
                    _collapsed = !_collapsed;

                    if (_collapsed)
                            _image.source = new imageExpander() as BitmapAsset;
                    else
                            _image.source = new imageCollapser() as BitmapAsset;
            }
      }
}
// END FILE "/VMap/src/com/vynance/controls/Collapser.as"

// BEGIN FILE "/VMap/src/com/vynance/controls/GlyphSignalGraph.as"
package com.vynance.controls
{
      import com.vynance.app.AppManager;
      import com.vynance.model.Ticker;
      import com.vynance.modules.Constants;

      import mx.core.UIComponent;

      public class GlyphSignalGraph extends UIComponent
      {
            private var _ticker:Ticker;

            override protected function
updateDisplayList(unscaledWidth:Number, unscaledHeight:Number):void
            {
                    var index:int;

                    this.graphics.clear();

                    this.graphics.beginFill(Constants.PANEL_BACKGROUND_COLOR);
                    this.graphics.drawRect(0, 0, unscaledWidth,
unscaledHeight);

                    if (_ticker == null)
                            return;

                    this.graphics.lineStyle(1, 0x888888);
                    this.graphics.moveTo(0, 0);
                    this.graphics.lineTo(0, unscaledHeight);
                    this.graphics.moveTo(0, (unscaledHeight - 1) / 2 );
                    this.graphics.lineTo(unscaledWidth, (unscaledHeight - 1) /
2);
```

```actionscript
                    this.graphics.moveTo(unscaledWidth - 1, 0);
                    this.graphics.lineTo(unscaledWidth - 1, unscaledHeight);

                    if (isNaN(_ticker.signal))
                            return;

                    this.graphics.lineStyle(1, 0x4444ff);

                    while (index <= AppManager.getInstance().signalGraphWidth)
{
                            if (_ticker.signalHistory[index] != null) {
                                    this.graphics.moveTo(index + 1,
getY(unscaledHeight, _ticker.signalHistory[index]));
                                    break;
                            }
                            index++;
                    }

                    while (index <= AppManager.getInstance().signalGraphWidth)
{
                            if (_ticker.signalHistory[index] != null)
                                    this.graphics.lineTo(index + 1,
getY(unscaledHeight, _ticker.signalHistory[index]));
                            index++;
                    }
            }

            public function setTicker(ticker:Ticker):void
            {
                    _ticker = ticker;
            }

            private function getY(unscaledHeight:Number, signal:Number):int
            {
                    const MAX_SIGNAL:int = 5;
                    var returnValue:int;

                    returnValue = (unscaledHeight + 1) / 2 - signal *
((unscaledHeight - 1) / 2) / MAX_SIGNAL;
                    if (returnValue < 0)
                        returnValue = 0;
                    if  (returnValue > unscaledHeight)
                            returnValue = unscaledHeight;
                    return returnValue;
            }
        }
}
// END FILE "/VMap/src/com/vynance/controls/GlyphSignalGraph.as"

// BEGIN FILE "/VMap/src/com/vynance/controls/GlyphSignalIndicator.as"
package com.vynance.controls
{
        import com.vynance.modules.Constants;

        import mx.core.UIComponent;
        import mx.utils.ColorUtil;
```

```actionscript
    public class GlyphSignalIndicator extends UIComponent
    {
        private var _color:int;

        public function set color(value:int):void
        {
            _color = value;
        }

        override protected function
updateDisplayList(unscaledWidth:Number, unscaledHeight:Number):void
        {
            this.graphics.clear();

            this.graphics.beginFill(_color);
            this.graphics.drawRoundRectComplex(0, 3.5, unscaledWidth,
unscaledHeight - 3.5, 0, 0, 3, 3);

            this.graphics.beginFill(ColorUtil.adjustBrightness(_color,
Constants.SIGNAL_LIGHT_COLOR_BRIGHTNESS));
            this.graphics.drawRoundRectComplex(0, 0, unscaledWidth,
3.5, 3, 3, 0, 0);

            this.graphics.endFill();
        }
    }
}
// END FILE "/VMap/src/com/vynance/controls/GlyphSignalIndicator.as"

// BEGIN FILE "/VMap/src/com/vynance/controls/GlyphSignalScale.as"
package com.vynance.controls
{
    import com.vynance.modules.Constants;

    import mx.core.UIComponent;
    import mx.utils.ColorUtil;

    public class GlyphSignalScale extends UIComponent
    {
        override protected function
updateDisplayList(unscaledWidth:Number, unscaledHeight:Number):void
        {
            const LIGHT_COLOR_HEIGHT:int = 2;
            const ROUNDED_CORNER_RADIUS:int = 2;

            this.graphics.clear();

    this.graphics.beginFill(ColorUtil.adjustBrightness(Constants.SIGNAL_COL
OR_0_TO_1, Constants.SIGNAL_LIGHT_COLOR_BRIGHTNESS));
            this.graphics.drawRoundRectComplex(0, 0, unscaledWidth / 6,
LIGHT_COLOR_HEIGHT, ROUNDED_CORNER_RADIUS, 0, 0, 0);
            this.graphics.beginFill(Constants.SIGNAL_COLOR_0_TO_1);
            this.graphics.drawRoundRectComplex(0, LIGHT_COLOR_HEIGHT,
unscaledWidth / 6, unscaledHeight - LIGHT_COLOR_HEIGHT, 0, 0,
ROUNDED_CORNER_RADIUS, 0);
```

```
            this.graphics.beginFill(ColorUtil.adjustBrightness(Constants.SIGNAL_COL
OR_1_TO_2, Constants.SIGNAL_LIGHT_COLOR_BRIGHTNESS));
                    this.graphics.drawRect(unscaledWidth / 6, 0, unscaledWidth
/ 6, LIGHT_COLOR_HEIGHT);
                    this.graphics.beginFill(Constants.SIGNAL_COLOR_1_TO_2);
                    this.graphics.drawRect(unscaledWidth / 6,
LIGHT_COLOR_HEIGHT, unscaledWidth / 6, unscaledHeight - LIGHT_COLOR_HEIGHT);

            this.graphics.beginFill(ColorUtil.adjustBrightness(Constants.SIGNAL_COL
OR_2_TO_3, Constants.SIGNAL_LIGHT_COLOR_BRIGHTNESS));
                    this.graphics.drawRect(unscaledWidth / 6 * 2, 0,
unscaledWidth / 6, LIGHT_COLOR_HEIGHT);
                    this.graphics.beginFill(Constants.SIGNAL_COLOR_2_TO_3);
                    this.graphics.drawRect(unscaledWidth / 6 * 2,
LIGHT_COLOR_HEIGHT, unscaledWidth / 6, unscaledHeight - LIGHT_COLOR_HEIGHT);

            this.graphics.beginFill(ColorUtil.adjustBrightness(Constants.SIGNAL_COL
OR_3_TO_4, Constants.SIGNAL_LIGHT_COLOR_BRIGHTNESS));
                    this.graphics.drawRect(unscaledWidth / 6 * 3, 0,
unscaledWidth / 6, LIGHT_COLOR_HEIGHT);
                    this.graphics.beginFill(Constants.SIGNAL_COLOR_3_TO_4);
                    this.graphics.drawRect(unscaledWidth / 6 * 3,
LIGHT_COLOR_HEIGHT, unscaledWidth / 6, unscaledHeight - LIGHT_COLOR_HEIGHT);

            this.graphics.beginFill(ColorUtil.adjustBrightness(Constants.SIGNAL_COL
OR_4_TO_5, Constants.SIGNAL_LIGHT_COLOR_BRIGHTNESS));
                    this.graphics.drawRect(unscaledWidth / 6 * 4, 0,
unscaledWidth / 6, LIGHT_COLOR_HEIGHT);
                    this.graphics.beginFill(Constants.SIGNAL_COLOR_4_TO_5);
                    this.graphics.drawRect(unscaledWidth / 6 * 4,
LIGHT_COLOR_HEIGHT, unscaledWidth / 6, unscaledHeight - LIGHT_COLOR_HEIGHT);

            this.graphics.beginFill(ColorUtil.adjustBrightness(Constants.SIGNAL_COL
OR_5_PLUS, Constants.SIGNAL_LIGHT_COLOR_BRIGHTNESS));
                    this.graphics.drawRoundRectComplex(unscaledWidth / 6 * 5,
0, unscaledWidth / 6, LIGHT_COLOR_HEIGHT, 0, ROUNDED_CORNER_RADIUS, 0, 0);
                    this.graphics.beginFill(Constants.SIGNAL_COLOR_5_PLUS);
                    this.graphics.drawRoundRectComplex(unscaledWidth / 6 * 5,
LIGHT_COLOR_HEIGHT, unscaledWidth / 6, unscaledHeight - LIGHT_COLOR_HEIGHT,
0, 0, 0, ROUNDED_CORNER_RADIUS);
                }
            }
}
// END FILE "/VMap/src/com/vynance/controls/GlyphSignalScale.as"

// BEGIN FILE "/VMap/src/com/vynance/controls/GlyphTicker.as"
package com.vynance.controls
{
        import com.vynance.app.AppManager;
        import com.vynance.model.Ticker;
        import com.vynance.modules.Constants;
        import com.vynance.modules.EnumMarketStatuses;
```

```actionscript
import com.vynance.utils.Signal;

import flash.events.MouseEvent;
import flash.text.TextField;
import flash.text.TextFormat;
import flash.utils.clearInterval;
import flash.utils.setInterval;

import mx.core.UIComponent;
import mx.utils.ColorUtil;

public class GlyphTicker extends UIComponent
{
    public static var format:TextFormat;

    public static const GLYPH_TICKER_HEIGHT:int = 17;

    public static const GLYPH_TICKER_WIDTH:int = 43;

    private var _hovered:Boolean;

    private var _intervalID:uint;

    private var _label:TextField;

    private var _tickerIndex:int;

    public function GlyphTicker()
    {
        super();

        if (format == null) {
            format = new TextFormat();
            format.align = "center";
            format.bold = true;
            format.color = 0xffffff;
            format.font = "Arial";
            format.size = 10;
        }

        _label = new TextField;
        _label.defaultTextFormat = format;
        _label.height = 17;
        _label.selectable = false;
        _label.visible = true;
        _label.width = GlyphTicker.GLYPH_TICKER_WIDTH;
        _label.x = 0;
        _label.y = 0;
        this.addChild(_label);

        this.addEventListener(MouseEvent.ROLL_OUT,
handle_ROLL_OUT);
        this.addEventListener(MouseEvent.ROLL_OVER,
handle_ROLL_OVER);
    }

    public function get tickerIndex():int
```

```actionscript
            {
                return _tickerIndex;
            }

            public function set tickerIndex(value:int):void
            {
                _tickerIndex = value;
            }

            override protected function
updateDisplayList(unscaledWidth:Number, unscaledHeight:Number):void
            {
                var color:int;
                var ticker:Ticker;

                if (_tickerIndex == -1) return;

                ticker =
AppManager.getInstance().map.tickerByIndex(_tickerIndex);

                this.graphics.clear();

                _label.text = ticker.symbol;

                this.graphics.beginFill(0, 0);
                this.graphics.drawRect(0, 0, unscaledWidth,
unscaledHeight);
                this.graphics.endFill();

                if (_hovered)
                    this.graphics.beginFill(0x2eaeff);
                else
                    this.graphics.beginFill(0x404040);
                this.graphics.drawRoundRect(1, 1,
GlyphTicker.GLYPH_TICKER_WIDTH - 2, GlyphTicker.GLYPH_TICKER_HEIGHT - 2, 6,
6);
                this.graphics.endFill();

                color = Signal.getColor(ticker);

                this.graphics.beginFill(color);
                this.graphics.drawRoundRectComplex(2, 5.5,
GlyphTicker.GLYPH_TICKER_WIDTH - 4, GlyphTicker.GLYPH_TICKER_HEIGHT - 7.5, 0,
0, 3, 3);
                this.graphics.endFill();

                this.graphics.beginFill(ColorUtil.adjustBrightness(color,
Constants.SIGNAL_LIGHT_COLOR_BRIGHTNESS));
                this.graphics.drawRoundRectComplex(2, 2,
GlyphTicker.GLYPH_TICKER_WIDTH - 4, 3.5, 3, 3, 0, 0);
                this.graphics.endFill();
            }

            public function updateFlashing():void
            {
                var ticker:Ticker;
```

```actionscript
                if (_tickerIndex == -1) return;

                ticker =
AppManager.getInstance().map.tickerByIndex(_tickerIndex);

                if (_intervalID != 0)
                    clearInterval(_intervalID);

                if (AppManager.getInstance().marketStatus ==
EnumMarketStatuses.CLOSED) {
                    _intervalID = 0;
                    this.alpha = 1;
                }
                else if (ticker.signalDeltaIndex <
AppManager.getInstance().flashingRank)
                    _intervalID = setInterval(Flash, 250);
                else {
                    _intervalID = 0;
                    this.alpha = 1;
                }
            }

        private function handle_ROLL_OUT(event:MouseEvent):void
        {
            _hovered = false;
            this.invalidateDisplayList();
            this.dispatchEvent(new
GlyphTickerEvent(GlyphTickerEvent.ROLL_OUT, _tickerIndex));
        }

        private function handle_ROLL_OVER(event:MouseEvent):void
        {
            _hovered = true;
            this.invalidateDisplayList();
            this.dispatchEvent(new
GlyphTickerEvent(GlyphTickerEvent.ROLL_OVER, _tickerIndex));
        }

        private function Flash():void {
            if (this.alpha == 1)
                this.alpha = 0.65;
            else
                this.alpha = 1;
        }
    }
}
// END FILE "/VMap/src/com/vynance/controls/GlyphTicker.as"

// BEGIN FILE "/VMap/src/com/vynance/controls/GlyphTickerEvent.as"
package com.vynance.controls
{
    import com.vynance.model.Ticker;

    import flash.events.Event;

    public class GlyphTickerEvent extends Event
```

```actionscript
        {
                public static const ROLL_OUT:String = "ROLL_OUT";

                public static const ROLL_OVER:String = "ROLL_OVER";

                private var _tickerIndex:int;

                public function GlyphTickerEvent(type:String, tickerIndex:int)
                {
                        super(type, true);

                        _tickerIndex = tickerIndex;
                }

                public function get tickerIndex():int
                {
                        return _tickerIndex;
                }
        }
}
// END FILE "/VMap/src/com/vynance/controls/GlyphTickerEvent.as"

// BEGIN FILE "/VMap/src/com/vynance/controls/Matrix.as"
package com.vynance.controls
{
        import com.vynance.app.AppManager;
        import com.vynance.model.Map;
        import com.vynance.model.MapEvent;
        import com.vynance.model.ParamLiquidity;
        import com.vynance.model.ParamMarketCap;
        import com.vynance.model.Ticker;
        import com.vynance.modules.EnumMarketStatuses;
        import com.vynance.modules.EnumTierTypes;

        import flash.display.DisplayObject;
        import flash.events.Event;
        import flash.text.TextField;
        import flash.text.TextFormat;

        import mx.controls.Label;
        import mx.core.UIComponent;

        public class Matrix extends UIComponent
        {
                private const COL_HEADER_HEIGHT:int = 27;

                private const GAP_H:int = 5;

                private const GAP_RIGHT:int = 25;

                private const GAP_V:int = 5;

                private const MATRIX_MIN_HEIGHT:int = 290;

                private const MATRIX_MIN_WIDTH:int = 621;

                private const ROW_HEADER_WIDTH:int = 81;
```

```actionscript
        private var _currentGlyph:int;

        private var _formatHeader:TextFormat;

        private var _formatTicker:TextFormat;

        private var _headerLabels:Array;

        private var _height:int;

        private var _labelUpdating:Label;

        private var _tickerGlyphs:Array;

        private var _width:int;

        public function Matrix()
        {
                _headerLabels = new Array();

                _formatHeader = new TextFormat("Verdana", 10, 0xff9000,
true);

                _labelUpdating = new Label();
                _labelUpdating.height = 17;
                _labelUpdating.selectable = false;
                _labelUpdating.setStyle("color", "#888888");
                _labelUpdating.text= "Updating...";
                _labelUpdating.width = 100;
                _labelUpdating.visible = false;
                this.addChild(_labelUpdating);

                for (var i:int = 0; i < 21; i++) {
                        var label:TextField = new TextField();
                        label.height = 17;
                        label.selectable = false;
                        _headerLabels.push(label);
                        this.addChild(label as DisplayObject);
                }

                _tickerGlyphs = new Array();

        AppManager.getInstance().map.addEventListener(MapEvent.BEGIN_SIGNAL_DOW
NLOAD, handle_Map_BeginSignalDownload);

        AppManager.getInstance().map.addEventListener(MapEvent.END_SIGNAL_DOWNL
OAD, handle_Map_EndSignalDownload);
        }

        override protected function commitProperties():void
        {
                if (this.height == 0)
                        return;

                var bucketHeight:int;
                var bucketWidth:int;
```

```
            var columnDimension:int;
            var currentHeaderLabel:int = 0;
            var glyph:GlyphTicker;
            var i:int;
            var label:TextField;
            var map:Map = AppManager.getInstance().map;
            var numberOfColumns:int;
            var numberOfRows:int;
            var rowDimension:int;

            if (map.tickerCount() == 0)
                 return;

            _height = this.height;
            if (_height < MATRIX_MIN_HEIGHT)
                 _height = MATRIX_MIN_HEIGHT;
            _width = this.width;
            if (_width < MATRIX_MIN_WIDTH)
                 _width = MATRIX_MIN_WIDTH;

            numberOfColumns = AppManager.getInstance().matrixColumns;
            numberOfRows = AppManager.getInstance().matrixRows;

            bucketHeight = (_height - COL_HEADER_HEIGHT - GAP_H *
(numberOfRows - 1)) / numberOfRows;
            bucketWidth = (_width - ROW_HEADER_WIDTH - GAP_RIGHT -
GAP_V * (numberOfColumns - 1)) / numberOfColumns;

            _currentGlyph = 0;

            if ((AppManager.getInstance().paramLiquidity.tierType ==
EnumTierTypes.Rows && AppManager.getInstance().paramMarketCap.tierType ==
EnumTierTypes.Columns) ||
                  (AppManager.getInstance().paramMarketCap.tierType ==
EnumTierTypes.Rows && AppManager.getInstance().paramLiquidity.tierType ==
EnumTierTypes.Columns)) {
                  label = TextField(_headerLabels[currentHeaderLabel]);
                  _formatHeader.align = "right";
                  label.defaultTextFormat = _formatHeader;
                  label.text = "0";
                  label.visible = true;
                  label.width = ROW_HEADER_WIDTH;
                  label.x = 0;
                  label.y = 10;
                  currentHeaderLabel++;
            }
            else if (AppManager.getInstance().paramLiquidity.tierType
== EnumTierTypes.Rows || AppManager.getInstance().paramMarketCap.tierType ==
EnumTierTypes.Rows) {
                  label = TextField(_headerLabels[currentHeaderLabel]);
                  _formatHeader.align = "right";
                  label.defaultTextFormat = _formatHeader;
                  label.text = "0";
                  label.visible = true;
                  label.width = ROW_HEADER_WIDTH;
                  label.x = 0;
                  label.y = 16;
```

```
                        currentHeaderLabel++;
                }
                else if (AppManager.getInstance().paramLiquidity.tierType
== EnumTierTypes.Columns || AppManager.getInstance().paramMarketCap.tierType
== EnumTierTypes.Columns) {
                        label = TextField(_headerLabels[currentHeaderLabel]);
                        _formatHeader.align = "left";
                        label.defaultTextFormat = _formatHeader;
                        label.text = "0";
                        label.visible = true;
                        label.width = ROW_HEADER_WIDTH;
                        label.x = ROW_HEADER_WIDTH - 6;
                        label.y = 10;
                        currentHeaderLabel++;
                }

                if (AppManager.DEBUG) {
                        AppManager.getInstance().log = "";
                        AppManager.getInstance().log +=
("blk_idx,gl_idx,tkr_idx,symbol,signal" + "\n");
                }

                for (var row:int = 0; row < numberOfRows; row++) {

                        var yCoor:int = COL_HEADER_HEIGHT + row *
(bucketHeight + GAP_H);

                        label = TextField(_headerLabels[currentHeaderLabel]);
                        _formatHeader.align = "right";
                        label.defaultTextFormat = _formatHeader;
                        label.width = ROW_HEADER_WIDTH;
                        label.x = 0;

                        if (AppManager.getInstance().paramCluster.tierType ==
EnumTierTypes.Rows || AppManager.getInstance().paramExchange.tierType ==
EnumTierTypes.Rows) {
                                if
(AppManager.getInstance().paramCluster.tierType == EnumTierTypes.Rows)
                                        label.text =
AppManager.getInstance().paramCluster.getText(row);
                                else
                                        label.text =
AppManager.getInstance().paramExchange.getText(row);
                                label.visible = true;
                                label.y = yCoor + bucketHeight / 2 - 17 / 2;
                        } else if
(AppManager.getInstance().paramLiquidity.tierType == EnumTierTypes.Rows ||
AppManager.getInstance().paramMarketCap.tierType == EnumTierTypes.Rows) {
                                if
(AppManager.getInstance().paramLiquidity.tierType == EnumTierTypes.Rows)
                                        label.text =
ParamLiquidity.formatLiquidity(AppManager.getInstance().paramLiquidity.tierMa
rkers[row]);
                                else
```

```
                                                label.text =
ParamMarketCap.formatMarketCap(AppManager.getInstance().paramMarketCap.tierMa
rkers[row]);
                                                label.visible = true;
                                                label.y = yCoor + bucketHeight + GAP_H / 2 - 17
/ 2;
                                        }
                                        else {
                                                label.visible = false;
                                        }

                                        currentHeaderLabel++;

                                        for (var column:int = 0; column < numberOfColumns;
column++) {

                                                var xCoor:int = ROW_HEADER_WIDTH + column *
(bucketWidth + GAP_V);

                                                if (row == 0) {

                                                        label =
TextField(_headerLabels[currentHeaderLabel]);
                                                        _formatHeader.align = "center";
                                                        label.defaultTextFormat = _formatHeader;
                                                        label.height = 17;
                                                        label.width = bucketWidth;
                                                        label.y = 10;

                                                        if
(AppManager.getInstance().paramCluster.tierType == EnumTierTypes.Columns ||
AppManager.getInstance().paramExchange.tierType == EnumTierTypes.Columns) {
                                                                if
(AppManager.getInstance().paramCluster.tierType == EnumTierTypes.Columns)
                                                                        label.text =
AppManager.getInstance().paramCluster.getText(column);
                                                                else
                                                                        label.text =
AppManager.getInstance().paramExchange.getText(column);
                                                                label.visible = true;
                                                                label.x = xCoor;
                                                        }
                                                        else if
(AppManager.getInstance().paramLiquidity.tierType == EnumTierTypes.Columns ||
AppManager.getInstance().paramMarketCap.tierType == EnumTierTypes.Columns) {
                                                                if
(AppManager.getInstance().paramLiquidity.tierType == EnumTierTypes.Columns)
                                                                        label.text =
ParamLiquidity.formatLiquidity(AppManager.getInstance().paramLiquidity.tierMa
rkers[column]);
                                                                else
                                                                        label.text =
ParamMarketCap.formatMarketCap(AppManager.getInstance().paramMarketCap.tierMa
rkers[column]);
                                                                label.visible = true;
                                                                label.x = xCoor + bucketWidth / 2 +
2;
```

```actionscript
                                }
                                else {
                                        label.visible = false;
                                }

                                currentHeaderLabel++;

                        }
                    }
                }

                for (i = currentHeaderLabel; i < _headerLabels.length; i++)
{
                        TextField(_headerLabels[i]).visible = false;
                }

                for (i = 0; i < _tickerGlyphs.length; i++) {
                        glyph = GlyphTicker(_tickerGlyphs[i]);
                        glyph.tickerIndex = -1;
                        glyph.visible = false;
                        glyph.y = -100;
                }

                arrangeBuckets();

                if (AppManager.DEBUG) {
                        AppManager.getInstance().log +=
("gl_idx,visible,tkr_idx,symbol,signal" + "\n");
                        for (i = 0; i < _tickerGlyphs.length; i++) {
                                glyph = GlyphTicker(_tickerGlyphs[i]);
                                AppManager.getInstance().log += (i.toString() +
",");
                                AppManager.getInstance().log +=
(glyph.visible.toString() + ",")
                                if (glyph.visible) {
                                        AppManager.getInstance().log +=
(glyph.tickerIndex.toString() + ",")
                                        AppManager.getInstance().log +=
(map.tickerByIndex(glyph.tickerIndex).symbol + ",")
                                        AppManager.getInstance().log +=
(map.tickerByIndex(glyph.tickerIndex).signal)
                                }
                                AppManager.getInstance().log += "\n";
                        }
                }

                for (i = 0; i < _tickerGlyphs.length; i++) {
                        GlyphTicker(_tickerGlyphs[i]).updateFlashing();
                }
            }

        override protected function
updateDisplayList(unscaledWidth:Number, unscaledHeight:Number):void
                {
                        var bucketHeight:int;
                        var bucketWidth:int;
                        var matrixOfColumns:int;
```

```actionscript
            var matrixOfRows:int;

            this.graphics.clear();

            matrixOfColumns = AppManager.getInstance().matrixColumns;
            matrixOfRows = AppManager.getInstance().matrixRows;

            bucketHeight = (_height - COL_HEADER_HEIGHT - GAP_H *
(matrixOfRows - 1)) / matrixOfRows;
            bucketWidth = (_width - ROW_HEADER_WIDTH - GAP_RIGHT -
GAP_V * (matrixOfColumns - 1)) / matrixOfColumns;

            for (var row:int = 0; row < matrixOfRows; row++) {

                var yCoor:int = COL_HEADER_HEIGHT + row *
(bucketHeight + GAP_H);

                for (var column:int = 0; column < matrixOfColumns;
column++) {

                    var xCoor:int = ROW_HEADER_WIDTH + column *
(bucketWidth + GAP_V);

                    this.graphics.beginFill(0x404040);
                    this.graphics.drawRoundRect(xCoor, yCoor,
bucketWidth, bucketHeight, 6, 6);

                    this.graphics.beginFill(0x292929);
                    this.graphics.drawRoundRect(xCoor + 1, yCoor +
1, bucketWidth - 2, bucketHeight - 2, 6, 6);

                    this.graphics.endFill();
                }
            }
        }

        private function handle_Map_BeginSignalDownload(event:Event):void
        {
            if (AppManager.getInstance().marketStatus ==
EnumMarketStatuses.OPEN)
                _labelUpdating.visible = true;
        }

        private function handle_Map_EndSignalDownload(event:Event):void
        {
            _labelUpdating.visible = false;

            callLater(this.invalidateProperties);
            callLater(this.invalidateDisplayList);
        }

        private function arrangeBuckets():void
        {
            const MARGIN_BOTTOM:int = 2;
            const MARGIN_LEFT:int = 2;
            const MARGIN_RIGHT:int = 2;
            const MARGIN_TOP:int = 2;
```

```actionscript
                var bucketHeight:int;
                var bucketWidth:int;
                var columnsInEachBucket:int;
                var currentColumn:Array = new Array();
                var currentGlyphIndex:int;
                var currentRow:Array = new Array();
                var matrixColumns:int;
                var matrixRows:int;
                var rowsInEachBucket:int;
                var signalIndex:int;
                var ticker:Ticker;
                var tickerGlyph:GlyphTicker;
                var tickerIndex:int;

                matrixRows = AppManager.getInstance().matrixRows;
                matrixColumns = AppManager.getInstance().matrixColumns;

                for (var bucket:int = 0; bucket < matrixRows *
matrixColumns; bucket++) {
                        currentColumn[bucket] = 0;
                        currentRow[bucket] = 0;
                }

                bucketHeight = (_height - COL_HEADER_HEIGHT - GAP_H *
(matrixRows - 1)) / matrixRows;
                bucketWidth = (_width - ROW_HEADER_WIDTH - GAP_RIGHT -
GAP_V * (matrixColumns - 1)) / matrixColumns;

                columnsInEachBucket = (bucketWidth - MARGIN_LEFT -
MARGIN_RIGHT) / GlyphTicker.GLYPH_TICKER_WIDTH;
                rowsInEachBucket = (bucketHeight - MARGIN_TOP -
MARGIN_BOTTOM) / GlyphTicker.GLYPH_TICKER_HEIGHT;

                for (signalIndex =
AppManager.getInstance().map.tickerCount() - 1; signalIndex >= 0;
signalIndex--) {

                        tickerIndex =
AppManager.getInstance().map.tickerIndexBySignalIndex(signalIndex);

                        ticker =
AppManager.getInstance().map.tickerByIndex(tickerIndex);

                        if (!ticker.excluded) {
                                if (currentRow[ticker.bucketIndex] <
rowsInEachBucket) {

                                        var xCoor:int = ROW_HEADER_WIDTH +
ticker.columnIndex * (bucketWidth + GAP_V) + MARGIN_LEFT +
GlyphTicker.GLYPH_TICKER_WIDTH * currentColumn[ticker.bucketIndex];
                                        var yCoor:int = COL_HEADER_HEIGHT +
ticker.rowIndex * (bucketHeight + GAP_H) + MARGIN_TOP +
GlyphTicker.GLYPH_TICKER_HEIGHT * currentRow[ticker.bucketIndex];

                                        if (currentGlyphIndex <
_tickerGlyphs.length)
```

```
                                        tickerGlyph =
_tickerGlyphs[currentGlyphIndex];
                                        else {
                                                tickerGlyph = new GlyphTicker();
                                                _tickerGlyphs.push(tickerGlyph);
                                                this.addChild(tickerGlyph);
                                        }
                                        currentGlyphIndex++;

                                        tickerGlyph.height =
GlyphTicker.GLYPH_TICKER_HEIGHT;
                                        tickerGlyph.tickerIndex = tickerIndex;
                                        tickerGlyph.visible = true;
                                        tickerGlyph.width =
GlyphTicker.GLYPH_TICKER_WIDTH;
                                        tickerGlyph.x = xCoor;
                                        tickerGlyph.y = yCoor;
                                        tickerGlyph.invalidateDisplayList();

                                        currentColumn[ticker.bucketIndex]++;
                                        if (currentColumn[ticker.bucketIndex] ==
columnsInEachBucket) {
                                                currentColumn[ticker.bucketIndex] =
0
                                                currentRow[ticker.bucketIndex]++;
                                        }
                                }
                        }
                }
        }
}
// END FILE "/VMap/src/com/vynance/controls/Matrix.as"

// BEGIN FILE "/VMap/src/com/vynance/model/Map.as"
package com.vynance.model
{
        import com.vynance.app.*;
        import com.vynance.modules.*;
        import com.vynance.utils.Signal;

        import flash.events.*;
        import flash.utils.Timer;

        public class Map extends EventDispatcher
        {
                private const URI_LOCATION:String =
"http://www.{domain}.com/{directory}/"
// COMMENT: HERE {domain} AND {directory} ARE PLACEHOLDERS TO BE MODIFIED

                private var _minutesSinceStartOfTrading:int;

                private var _orderedByLiquidity:Array;

                private var _orderedByMarketCap:Array;

                private var _signalIndexes:Array;
```

```actionscript
        private var _tickers:Array;

        private var _timer:Timer;

        private var _timerMap:Timer;

        private var _urlDownloaderMap:URLDownloader;

        private var _urlDownloaderSignal:URLDownloader;

        private var _urlDownloaderTime:URLDownloader;

        public function Map()
        {
                _timer = new Timer(30000);
                _timer.addEventListener(TimerEvent.TIMER, handle_timer);

                _urlDownloaderMap = new URLDownloader();

        _urlDownloaderMap.addEventListener(URLDownloaderEvent.COMPLETE,
handle_MapDownloaded);

                _urlDownloaderSignal = new URLDownloader();

        _urlDownloaderSignal.addEventListener(URLDownloaderEvent.COMPLETE,
handle_SignalDownloaded);

                _urlDownloaderTime = new URLDownloader();

        _urlDownloaderTime.addEventListener(URLDownloaderEvent.COMPLETE,
handle_TimeDownloaded);
        }

        public function get maxLiquidity():Number
        {
                return
Ticker(_tickers[_orderedByLiquidity[_orderedByLiquidity.length -
1]]).liquidity;
        }

        public function get maxMarketCap():Number
        {
                return
Ticker(_tickers[_orderedByMarketCap[_orderedByMarketCap.length -
1]]).marketCap;
        }

        public function get minLiquidity():Number
        {
                return Ticker(_tickers[_orderedByLiquidity[0]]).liquidity;
        }

        public function get minMarketCap():Number
        {
                return Ticker(_tickers[_orderedByMarketCap[0]]).marketCap;
        }
```

```actionscript
public function get minutesSinceStartOfTrading():int
{
        return _minutesSinceStartOfTrading;
}

public function downloadMap():void
{
        _urlDownloaderMap.download(URI_LOCATION + "m.txt");
}

public function recalculate():void
{
        for (var i:int = 0; i < _tickers.length; i++) {
                Ticker(_tickers[i]).reset();
        }

        applyTiers();

        applyFilters();
}

public function tickerByIndex(tickerIndex:int):Ticker
{
        return _tickers[tickerIndex];
}

public function tickerIndexBySignalIndex(signalIndex:int):int
{
        return _signalIndexes[signalIndex];
}

public function tickerCount():int
{
        if (_tickers == null)
                return 0;
        else
                return _tickers.length;
}

private function handle_MapDownloaded(event:Event):void
{
        var html:String;
        var lines:Array;
        var ticker:Ticker;
        var values:Array;

        html = _urlDownloaderMap.urlLoader.data;

_urlDownloaderMap.removeEventListener(URLDownloaderEvent.COMPLETE,
handle_MapDownloaded);
        _urlDownloaderMap = null;

        lines = html.split("\n");

        _orderedByLiquidity = new Array(lines.length - 1);
```

```actionscript
                _orderedByMarketCap = new Array(lines.length - 1);
                _signalIndexes = new Array(lines.length - 1);
                _tickers = new Array(lines.length - 1);

                for (var i:int = 0; i < lines.length; i++) {

                        values = String(lines[i]).split("\t");

                        var symbol:String = values[0];
                        var cluster:int = values[1];
                        var exchange:int = values[2];
                        var marketCap:Number = values[3];
                        var marketCapIndex:int = values[4];
                        var liquidity:Number = values[5];
                        var liquidityIndex:int = values[6];

                        ticker = new Ticker(symbol, cluster, exchange,
liquidity, marketCap);

                        _tickers[i] = ticker;
                        _orderedByLiquidity[liquidityIndex] = i;
                        _orderedByMarketCap[marketCapIndex] = i;
                }

                this.dispatchEvent(new MapEvent(MapEvent.MAP_DOWNLOADED));

                _urlDownloaderSignal.download(URI_LOCATION + "s.txt");
        }

        private function handle_SignalDownloaded(event:Event):void
        {
                var index:int;
                var html:String;
                var multiplier:Number;
                var pos1:int;
                var pos2:int;
                var values:Array;

                html = _urlDownloaderSignal.urlLoader.data;

                pos2 = html.indexOf(",");
                _minutesSinceStartOfTrading = int(html.substr(0, pos2));

                while (true) {
                        pos1 = pos2 + 1;
                        pos2 = html.indexOf(",", pos1);
                        if (pos2 < 0)
                                break;
                        ParseSignalFileLine(html.substr(pos1, pos2 - pos1),
index);

                        index++;
                }
                ParseSignalFileLine( html.substr(pos1), index);

                this.recalculate();
```

```actionscript
                        this.dispatchEvent(new
MapEvent(MapEvent.END_SIGNAL_DOWNLOAD));

                        if (_minutesSinceStartOfTrading == -1) {
                                AppManager.getInstance().marketStatus =
EnumMarketStatuses.CLOSED;
                                this.dispatchEvent(new
MapEvent(MapEvent.MARKET_IS_CLOSED));
                                return;
                        }
                        else if (_minutesSinceStartOfTrading == 0) {
                                AppManager.getInstance().marketStatus =
EnumMarketStatuses.PRE_OPEN;
                        }
                        else{
                                AppManager.getInstance().marketStatus =
EnumMarketStatuses.OPEN;
                        }

                        _urlDownloaderTime.download(URI_LOCATION + "t.asp");
                }

            private function handle_TimeDownloaded(event:Event):void
            {
                        var delay:int;
                        var seconds:int;
                        var time:String;

                        if (AppManager.getInstance().marketStatus ==
EnumMarketStatuses.CLOSED)
                                return;

                        time = _urlDownloaderTime.urlLoader.data;
                        seconds = int(time.substr(time.length - 5, 2));

                        if (seconds < 12)
                                delay = 12 - seconds;
                        else if (seconds < 42)
                                delay = 42 - seconds;
                        else
                                delay = 72 - seconds;

                        if (delay < 5)
                                delay += 30

                        _timer.delay = delay * 1000;
                        _timer.start();
                }

            private function handle_timer(event:Event):void
            {
                        _timer.stop();

                        this.dispatchEvent(new
MapEvent(MapEvent.BEGIN_SIGNAL_DOWNLOAD));

                        _urlDownloaderSignal.download(URI_LOCATION + "s.txt");
```

```actionscript
        }

        private function handle_timerMap(event:Event):void
        {
                _timerMap.stop();
                _urlDownloaderMap.download(URI_LOCATION + "m.txt");
        }

        private function applyFilters():void
        {
                var i:int;
                var paramCluster:ParamCluster =
AppManager.getInstance().paramCluster;
                var paramExchange:ParamExchange=
AppManager.getInstance().paramExchange;
                var paramLiquidity:ParamLiquidity =
AppManager.getInstance().paramLiquidity;
                var paramMarketCap:ParamMarketCap =
AppManager.getInstance().paramMarketCap;
                var paramSignal:ParamSignal =
AppManager.getInstance().paramSignal;
                var ticker:Ticker;

                for (i = 0; i < _tickers.length; i++) {
                        ticker = Ticker(_tickers[i]);
                        if (!paramCluster.isClusterSelected(ticker.cluster))
                                ticker.excluded = true;
                }

                for (i = 0; i < _tickers.length; i++) {
                        ticker = Ticker(_tickers[i]);
                        if
(!paramExchange.isExchangeSelected(ticker.exchange))
                                ticker.excluded = true;
                }

                if (paramLiquidity.tierType == EnumTierTypes.None) {
                        for (i = 0; i < _tickers.length; i++) {
                                ticker = Ticker(_tickers[i])
                                if (ticker.liquidity < paramLiquidity.min)
                                        ticker.excluded = true;
                                if (ticker.liquidity > paramLiquidity.max)
                                        ticker.excluded = true;
                        }
                }

                if (paramMarketCap.tierType == EnumTierTypes.None) {
                        for (i = 0; i < _tickers.length; i++) {
                                ticker = Ticker(_tickers[i])
                                if (ticker.marketCap < paramMarketCap.min)
                                        ticker.excluded = true;
                                if (ticker.marketCap > paramMarketCap.max)
                                        ticker.excluded = true;
                        }
                }

                for (i = 0; i < _tickers.length; i++) {
```

```
                        ticker = Ticker(_tickers[i])
                        if (Signal.compareSignal(Math.abs(ticker.signal),
paramSignal.min) == -1)
                              ticker.excluded = true;
                        if (Signal.compareSignal(Math.abs(ticker.signal),
paramSignal.max) == 1)
                              ticker.excluded = true;
                        if (paramSignal.min    > 0) {
                              if (isNaN(ticker.signal))
                                    ticker.excluded = true;
                        }
                  }
            }

            private function applyTiers():void
            {
                  var i:int;
                  var paramCluster:ParamCluster =
AppManager.getInstance().paramCluster;
                  var paramExchange:ParamExchange=
AppManager.getInstance().paramExchange;
                  var paramMarketCap:ParamMarketCap =
AppManager.getInstance().paramMarketCap;
                  var tier:int;
                  var tierInRows:Boolean;
                  var upperBound:int;

                  AppManager.getInstance().matrixRows = 1;
                  AppManager.getInstance().matrixColumns = 1;

                  if (paramCluster.isTiered) {
                        tierInRows = (paramCluster.tierType ==
EnumTierTypes.Rows);
                        for (i = 0; i < _tickers.length; i++) {
                              if (tierInRows)
                                    Ticker(_tickers[i]).rowIndex =
paramCluster.getSelectedClusterIndex(Ticker(_tickers[i]).cluster);
                              else
                                    Ticker(_tickers[i]).columnIndex =
paramCluster.getSelectedClusterIndex(Ticker(_tickers[i]).cluster);
                        }
                        if (tierInRows)
                              AppManager.getInstance().matrixRows =
paramCluster.countOfSelectedClusters;
                        else
                              AppManager.getInstance().matrixColumns =
paramCluster.countOfSelectedClusters;
                  }

                  if (paramExchange.isTiered) {
                        tierInRows = (paramExchange.tierType ==
EnumTierTypes.Rows);
                        for (i = 0; i < _tickers.length; i++) {
                              if (tierInRows)
                                    Ticker(_tickers[i]).rowIndex =
paramExchange.getSelectedExchangeIndex(Ticker(_tickers[i]).exchange);
                              else
```

```
                                        Ticker(_tickers[i]).columnIndex =
paramExchange.getSelectedExchangeIndex(Ticker(_tickers[i]).exchange);
                        }
                        if (tierInRows)
                                AppManager.getInstance().matrixRows =
paramExchange.countOfSelectedExchanges;
                        else
                                AppManager.getInstance().matrixColumns =
paramExchange.countOfSelectedExchanges;
                }

                applyTierLiquidity();

                applyTierMarketCap();
        }

        private function applyTierLiquidity():void
        {
                var i:int;
                var maxIndex:int = -1;
                var minIndex:int = -1;
                var paramLiquidity:ParamLiquidity =
AppManager.getInstance().paramLiquidity;
                var ticker:Ticker;
                var tier:int = -1;
                var tierInRows:Boolean = (paramLiquidity.tierType ==
EnumTierTypes.Rows);
                var upperBound:int = -1;

                if (!paramLiquidity.isTiered) return;

                for (i = 0; i < _orderedByLiquidity.length; i++) {
                        ticker = Ticker(_tickers[_orderedByLiquidity[i]]);
                        if (ticker.liquidity > paramLiquidity.min) {
                                minIndex = i;
                                break;
                        }
                }

                for (i = _orderedByLiquidity.length - 1; i >= 0; i--) {
                        ticker = Ticker(_tickers[_orderedByLiquidity[i]]);
                        if (ticker.liquidity < paramLiquidity.max) {
                                maxIndex = i;
                                break;
                        }
                }

                paramLiquidity.emptyMarkers();

                for (i = 0; i < _orderedByLiquidity.length; i++) {
                        if (i == 0 || i > upperBound) {
                                tier++;
                                upperBound = _orderedByLiquidity.length * (tier
+ 1) / paramLiquidity.numberOfTiers - 1;

        paramLiquidity.tierMarkers.push(Ticker(_tickers[_orderedByLiquidity[upp
erBound]]).liquidity);
```

```actionscript
                    }
                    if (tierInRows)

        Ticker(_tickers[_orderedByLiquidity[i]]).rowIndex = tier;
                    else

        Ticker(_tickers[_orderedByLiquidity[i]]).columnIndex = tier;
                }

        paramLiquidity.tierMarkers.push(Ticker(_tickers[_orderedByLiquidity[_or
deredByLiquidity.length - 1]]).liquidity);

                    if (tierInRows)
                        AppManager.getInstance().matrixRows =
paramLiquidity.numberOfTiers;
                    else
                        AppManager.getInstance().matrixColumns =
paramLiquidity.numberOfTiers;
            }

        private function applyTierMarketCap():void
        {
                var i:int;
                var maxIndex:int = -1;
                var minIndex:int = -1;
                var paramMarketCap:ParamMarketCap =
AppManager.getInstance().paramMarketCap;
                var ticker:Ticker;
                var tier:int = -1;
                var tierInRows:Boolean = (paramMarketCap.tierType ==
EnumTierTypes.Rows);
                var upperBound:int = -1;

                if (!paramMarketCap.isTiered) return;

                for (i = 0; i < _orderedByMarketCap.length; i++) {
                    ticker = Ticker(_tickers[_orderedByMarketCap[i]]);
                    if (ticker.marketCap > paramMarketCap.min) {
                        minIndex = i;
                        break;
                    }
                }

                for (i = _orderedByMarketCap.length - 1; i >= 0; i--) {
                    ticker = Ticker(_tickers[_orderedByMarketCap[i]]);
                    if (ticker.marketCap < paramMarketCap.max) {
                        maxIndex = i;
                        break;
                    }
                }

                paramMarketCap.emptyMarkers();

                for (i = 0; i < _orderedByMarketCap.length; i++) {
                    if (i == 0 || i > upperBound) {
                        tier++;
```

```
                              upperBound = _orderedByMarketCap.length * (tier
+ 1) / paramMarketCap.numberOfTiers - 1;

     paramMarketCap.tierMarkers.push(Ticker(_tickers[_orderedByMarketCap[upp
erBound]]).marketCap);
                    }
                    if (tierInRows)

     Ticker(_tickers[_orderedByMarketCap[i]]).rowIndex = tier;
                    else

     Ticker(_tickers[_orderedByMarketCap[i]]).columnIndex = tier;
                }

     paramMarketCap.tierMarkers.push(Ticker(_tickers[_orderedByMarketCap[_or
deredByMarketCap.length - 1]]).marketCap);

                if (tierInRows)
                    AppManager.getInstance().matrixRows =
paramMarketCap.numberOfTiers;
                else
                    AppManager.getInstance().matrixColumns =
paramMarketCap.numberOfTiers;
            }

        private function DescrambleSignal(html:String, index:int):Number
        {
            var multiplier:Number;
            var signal:Number;

            if (html.length == 0)
                return NaN;

            multiplier = Math.sin(Math.sqrt(3) * (index + 2) +
Math.sqrt(7) * Math.cos(Math.sqrt(11) * (index + 2)));
            if (Math.round(multiplier * 100) == 0)
                multiplier = Math.cos(Math.sqrt(3) * (index + 2) +
Math.sqrt(7) * Math.sin(Math.sqrt(11) * (index + 2)));
            multiplier = Math.round(multiplier * 100) / 100;

            signal = Number(html) / multiplier;
            signal = Math.round(signal * 100) / 100;
            return signal;
        }

        private function DescrambleSignal2(html:String, index:int):Number
        {
            var multiplier:Number;
            var signal:Number;

            multiplier = Math.sin(Math.sqrt(3) * (index + 2) +
Math.sqrt(7) * Math.cos(Math.sqrt(11) * (index + 2)));
            signal = Number(html) / multiplier;
            signal = Math.round(signal * 100) / 100;

            return signal;
        }
```

```actionscript
            private function ParseSignalFileLine(line:String, index:int):void
            {
                var values:Array;

                values = line.split("\t");
                Ticker(_tickers[index]).signal =
DescrambleSignal(values[0], index);
                _signalIndexes[int(values[1])] = index;
                if (values.length == 3)
                    Ticker(_tickers[index]).signalDeltaIndex =
int(values[2]);
                else
                    Ticker(_tickers[index]).signalDeltaIndex = 9999;
            }
        }
}
// END FILE "/VMap/src/com/vynance/model/Map.as"

// BEGIN FILE "/VMap/src/com/vynance/model/MapEvent.as"
package com.vynance.model
{
    import flash.events.Event;

    public class MapEvent extends Event
        {
            public static const BEGIN_SIGNAL_DOWNLOAD:String =
"BEGIN_SIGNAL_DOWNLOAD";

            public static const END_SIGNAL_DOWNLOAD:String =
"END_SIGNAL_DOWNLOAD";

            public static const MAP_DOWNLOADED:String = "MAP_DOWNLOADED";

            public static const MARKET_IS_CLOSED:String = "MARKET_IS_CLOSED";

            public function MapEvent(type:String)
            {
                super(type, true);
            }
        }
}
// END FILE "/VMap/src/com/vynance/model/MapEvent.as"

// BEGIN FILE "/VMap/src/com/vynance/model/ParamCluster.as"
package com.vynance.model
{
    import com.vynance.modules.EnumClusters;
    import com.vynance.modules.EnumTierTypes;

    public class ParamCluster
        {
            public static function getClusterText(clusterID:int):String
            {
                switch (clusterID) {
                    case EnumClusters.CLUSTER0: return "Cyclicals";
                    case EnumClusters.CLUSTER1: return "Energy";
```

```actionscript
                case EnumClusters.CLUSTER2: return "Financials";
                case EnumClusters.CLUSTER3: return "Healthcare";
                case EnumClusters.CLUSTER4: return "Industrials";
                case EnumClusters.CLUSTER5: return "Materials";
                case EnumClusters.CLUSTER6: return "Non-Cyclicals";
                case EnumClusters.CLUSTER7: return "Technology";
                case EnumClusters.CLUSTER8: return "Telecom";
                case EnumClusters.CLUSTER9: return "Utilities";
        }
        return "INVALID CLUSTER";
    }

    private var _countOfSelectedClusters:int;

    private var _clusterPositions:Array;

    private var _tierType:int;

    public function ParamCluster()
    {
        _clusterPositions = new Array();

        _clusterPositions[EnumClusters.CLUSTER0] = 0;
        _clusterPositions[EnumClusters.CLUSTER1] = 1;
        _clusterPositions[EnumClusters.CLUSTER2] = 2;
        _clusterPositions[EnumClusters.CLUSTER3] = 3;
        _clusterPositions[EnumClusters.CLUSTER4] = 4;
        _clusterPositions[EnumClusters.CLUSTER5] = 5;
        _clusterPositions[EnumClusters.CLUSTER6] = 6;
        _clusterPositions[EnumClusters.CLUSTER7] = 7;
        _clusterPositions[EnumClusters.CLUSTER8] = 8;
        _clusterPositions[EnumClusters.CLUSTER9] = 9;

        _countOfSelectedClusters = _clusterPositions.length;
    }

    public function get countOfSelectedClusters():int
    {
        return _countOfSelectedClusters;
    }

    public function get isTiered():Boolean
    {
        return (_tierType != EnumTierTypes.None);
    }

    public function get tierType():int
    {
        return _tierType;
    }

    public function set tierType(value:int):void
    {
        _tierType = value;
    }

    public function addSelectedCluster(clusterID:int):void
```

```actionscript
                {
                        _clusterPositions[clusterID] = _countOfSelectedClusters;
                        _countOfSelectedClusters++;
                }

                public function clearSelectedClusters():void
                {
                        for (var i:int = 0; i < EnumClusters.COUNT; i++) {
                                _clusterPositions[i] = -1;
                        }
                        _countOfSelectedClusters = 0;
                }

                public function getSelectedClusterIndex(clusterID:int):int
                {
                        return _clusterPositions[clusterID];
                }

                public function getText(position:int):String
                {
                        for (var clusterID:int = 0; clusterID <
_clusterPositions.length; clusterID++) {
                                if (_clusterPositions[clusterID] > -1) {
                                        if (position == 0)
                                                return
ParamCluster.getClusterText(clusterID);
                                        position--;
                                }
                        }
                        return "Invalid cluster";
                }

                public function isClusterSelected(clusterID:int):Boolean
                {
                        return (_clusterPositions[clusterID] > -1);
                }
        }
}
// END FILE "/VMap/src/com/vynance/model/ParamCluster.as"

// BEGIN FILE "/VMap/src/com/vynance/model/ParamExchange.as"
package com.vynance.model
{
        import com.vynance.modules.EnumExchanges;
        import com.vynance.modules.EnumTierTypes;

        public class ParamExchange
        {
                public static function getExchangeText(exchangeID:int):String
                {
                        switch (exchangeID) {
                                case EnumExchanges.AMEX: return "AMEX";
                                case EnumExchanges.NSDQ: return "NASDAQ";
                                case EnumExchanges.NYSE: return "NYSE";
                        }
                        return "INVALID EXCHANGE";
                }
        }
```

```actionscript
private var _countOfSelectedExchanges:int;

private var _exchangePositions:Array;

private var _tierType:int;

public function ParamExchange()
{
        _exchangePositions = new Array();

        _exchangePositions[EnumExchanges.AMEX] = 0;
        _exchangePositions[EnumExchanges.NSDQ] = 1;
        _exchangePositions[EnumExchanges.NYSE] = 2;

        _countOfSelectedExchanges = _exchangePositions.length;
}

public function get countOfSelectedExchanges():int
{
        return _countOfSelectedExchanges;
}

public function get isTiered():Boolean
{
        return (_tierType != EnumTierTypes.None);
}

public function get tierType():int
{
        return _tierType;
}

public function set tierType(value:int):void
{
        _tierType = value;
}

public function addSelectedExchange(exchangeID:int):void
{
        _exchangePositions[exchangeID] = _countOfSelectedExchanges;
        _countOfSelectedExchanges++;
}

public function clearSelectedExchanges():void
{
        for (var i:int = 0; i < EnumExchanges.COUNT; i++) {
                _exchangePositions[i] = -1;
        }
        _countOfSelectedExchanges = 0;
}

public function getSelectedExchangeIndex(exchangeID:int):int
{
        return _exchangePositions[exchangeID];
}
```

```
            public function getText(position:int):String
            {
                    for (var exchangeID:int = 0; exchangeID <
_exchangePositions.length; exchangeID++) {
                            if (_exchangePositions[exchangeID] > -1) {
                                    if (position == 0) {
                                            return
ParamExchange.getExchangeText(exchangeID);
                                    }
                                    position--;
                            }
                    }
                    return "Invalid exchange";
            }

            public function isExchangeSelected(exchangeID:int):Boolean
            {
                    return (_exchangePositions[exchangeID] > -1);
            }
        }
}
// END FILE "/VMap/src/com/vynance/model/ParamExchange.as"

// BEGIN FILE "/VMap/src/com/vynance/model/ParamLiquidity.as"
package com.vynance.model
{
        import com.vynance.modules.EnumTierTypes;
        import mx.formatters.NumberFormatter;

        public class ParamLiquidity
        {
                private static var _formatter:NumberFormatter;

                public static function formatLiquidity(value:Number):String
                {
                        if (_formatter == null)
                                _formatter = new NumberFormatter();

                        if (value == 0)
                                return "N/A";

                        switch (value.toString().length) {
                                case 1:
                                        return value.toString();
                                case 2:
                                        return value.toString();
                                case 3:
                                        return value.toString();
                                case 4:
                                        _formatter.precision = 2;
                                        return _formatter.format(value / 1000) + " K";
                                case 5:
                                        _formatter.precision = 1;
                                        return _formatter.format(value / 1000) + " K";

                                case 6:
```

```actionscript
                                return Math.round(value / 1000).toString() + "
K";
                        case 7:
                                _formatter.precision = 2;
                                return _formatter.format(value / 1000000) + "
M";
                        case 8:
                                _formatter.precision = 1;
                                return _formatter.format(value / 1000000) + "
M";
                        case 9:
                                return Math.round(value / 1000000).toString() +
" M";
                        case 10:
                                _formatter.precision = 2;
                                return _formatter.format(value / 1000000000) +
" B";
                        case 11:
                                _formatter.precision = 1;
                                return _formatter.format(value / 1000000000) +
" B";
                        case 12:
                                return Math.round(value /
1000000000).toString() + " B";
                }
                return "Invalid liquidity";
        }

        private var _max:Number;

        private var _min:Number;

        private var _numberOfTiers:int;

        private var _tierMarkers:Array;

        private var _tierType:int;

        public function ParamLiquidity()
        {
                _max = Number.MAX_VALUE;
        }

        public function get isTiered():Boolean
        {
                return (_tierType != EnumTierTypes.None);
        }

        public function get max():Number
        {
                return _max;
        }

        public function set max(value:Number):void
        {
                _max = value;
        }
```

```actionscript
            public function get min():Number
            {
                    return _min;
            }

            public function set min(value:Number):void
            {
                    _min = value;
            }

            public function get numberOfTiers():int
            {
                    return _numberOfTiers;
            }

            public function set numberOfTiers(value:int):void
            {
                    _numberOfTiers = value;
            }

            public function get tierMarkers():Array
            {
                    return _tierMarkers;
            }

            public function get tierType():int
            {
                    return _tierType;
            }

            public function set tierType(value:int):void
            {
                    _tierType = value;
            }

            public function emptyMarkers():void
            {
                    _tierMarkers = new Array();
            }
        }
}
// END FILE "/VMap/src/com/vynance/model/ParamLiquidity.as"

// BEGIN FILE "/VMap/src/com/vynance/model/ParamMarketCap.as"
package com.vynance.model
{
        import com.vynance.modules.EnumTierTypes;
        import mx.formatters.NumberFormatter;

        public class ParamMarketCap
        {
                private static var _formatter:NumberFormatter;

                public static function formatMarketCap(value:Number):String
                {
                        if (_formatter == null)
```

```actionscript
                    _formatter = new NumberFormatter();

            if (value == 0)
                return "N/A";

            switch (value.toString().length) {
                case 1:
                    return value.toString();
                case 2:
                    return value.toString();
                case 3:
                    return value.toString();
                case 4:
                    _formatter.precision = 2;
                    return _formatter.format(value / 1000) + " K";
                case 5:
                    _formatter.precision = 1;
                    return _formatter.format(value / 1000) + " K";

                case 6:
                    return Math.round(value / 1000).toString() + "
K";

                case 7:
                    _formatter.precision = 2;
                    return _formatter.format(value / 1000000) + "
M";

                case 8:
                    _formatter.precision = 1;
                    return _formatter.format(value / 1000000) + "
M";

                case 9:
                    return Math.round(value / 1000000).toString() +
" M";

                case 10:
                    _formatter.precision = 2;
                    return _formatter.format(value / 1000000000) +
" B";

                case 11:
                    _formatter.precision = 1;
                    return _formatter.format(value / 1000000000) +
" B";

                case 12:
                    return Math.round(value /
1000000000).toString() + " B";
            }
            return "Invalid market cap";
        }

        private var _max:Number;

        private var _min:Number;

        private var _numberOfTiers:int;

        private var _tierMarkers:Array;

        private var _tierType:int;
```

```actionscript
public function ParamMarketCap()
{
        _max = Number.MAX_VALUE;
}

public function get isTiered():Boolean
{
        return (_tierType != EnumTierTypes.None);
}

public function get max():Number
{
        return _max;
}

public function set max(value:Number):void
{
        _max = value;
}

public function get min():Number
{
        return _min;
}

public function set min(value:Number):void
{
        _min = value;
}

public function get numberOfTiers():int
{
        return _numberOfTiers;
}

public function set numberOfTiers(value:int):void
{
        _numberOfTiers = value;
}

public function get tierMarkers():Array
{
        return _tierMarkers;
}

public function get tierType():int
{
        return _tierType;
}

public function set tierType(value:int):void
{
        _tierType = value;
}

public function emptyMarkers():void
```

```actionscript
                {
                    _tierMarkers = new Array();
                }
        }
}
// END FILE "/VMap/src/com/vynance/model/ParamMarketCap.as"

// BEGIN FILE "/VMap/src/com/vynance/model/ParamSignal.as"
package com.vynance.model
{
        public class ParamSignal
        {
                private var _max:Number;

                private var _min:Number;

                public function ParamSignal()
                {
                    _max = 9999;
                }

                public function get max():Number
                {
                    return _max;
                }

                public function set max(value:Number):void
                {
                    _max = value;
                }

                public function get min():Number
                {
                    return _min;
                }

                public function set min(value:Number):void
                {
                    _min = value;
                }
        }
}
// END FILE "/VMap/src/com/vynance/model/ParamSignal.as"

// BEGIN FILE "/VMap/src/com/vynance/model/Ticker.as"
package com.vynance.model
{
import com.vynance.app.AppManager;

        public class Ticker
        {
                public var columnIndex:int;

                public var excluded:Boolean;

                public var rowIndex:int;
```

```actionscript
        public var signalDeltaIndex:int;

        private var _cluster:int;

        private var _exchange:int;

        private var _liquidity:Number;

        private var _marketCap:Number;

        private var _signal:Number;

        private var _signalChange:Number;

        private var _symbol:String;

        public function Ticker(symbol:String, cluster:int, exchange:int,
liquidity:Number, marketCap:Number)
        {
                _symbol = symbol;
                _cluster = cluster;
                _exchange = exchange;
                _liquidity = liquidity;
                _marketCap = marketCap;

                _signal = NaN;
                _signalChange = NaN;
        }

        public function get bucketIndex():int
        {
                return this.rowIndex *
AppManager.getInstance().matrixColumns + this.columnIndex;
        }

        public function get cluster():int
        {
                return _cluster;
        }

        public function get exchange():int
        {
                return _exchange;
        }

        public function get liquidity():Number
        {
                return _liquidity;
        }

        public function get marketCap():Number
        {
                return _marketCap;
        }

        public function get signal():Number
        {
```

```actionscript
                return _signal;
        }

        public function set signal(value:Number):void
        {
                if (!isNaN(_signal))
                        _signalChange = value - _signal;

                _signal = value;
        }

        public function get signalChange():Number
        {
                return _signalChange;
        }

        public function get symbol():String
        {
                return _symbol;
        }

        public function reset():void
        {
                this.columnIndex = 0;
                this.excluded = false;
                this.rowIndex = 0;
        }
    }
}
// END FILE "/VMap/src/com/vynance/model/Ticker.as"

// BEGIN FILE "/VMap/src/com/vynance/modules/Constants.as"
package com.vynance.modules
{
        public class Constants
        {
                public static const PANEL_BACKGROUND_COLOR:int = 0x404040;

                public static const SIGNAL_COLOR_0_TO_1:int = 0x666666;
                public static const SIGNAL_COLOR_1_TO_2:int = 0x40B06C;
                public static const SIGNAL_COLOR_2_TO_3:int = 0x3380C2;
                public static const SIGNAL_COLOR_3_TO_4:int = 0xF4D701;
                public static const SIGNAL_COLOR_4_TO_5:int = 0xFF9C2C;
                public static const SIGNAL_COLOR_5_PLUS:int = 0xF53636;
                public static const SIGNAL_COLOR_NA:int = 0xB4B4B4;
                public static const SIGNAL_LIGHT_COLOR_BRIGHTNESS:int = 64;
        }
}
// END FILE "/VMap/src/com/vynance/modules/Constants.as"

// BEGIN FILE "/VMap/src/com/vynance/modules/EnumClusters.as"
package com.vynance.modules
{
        public class EnumClusters
        {
                public static const CLUSTER0:int = 0;
                public static const CLUSTER1:int = 1;
```

```actionscript
            public static const CLUSTER2:int = 2;
            public static const CLUSTER3:int = 3;
            public static const CLUSTER4:int = 4;
            public static const CLUSTER5:int = 5;
            public static const CLUSTER6:int = 6;
            public static const CLUSTER7:int = 7;
            public static const CLUSTER8:int = 8;
            public static const CLUSTER9:int = 9;
            public static const COUNT:int = 10;
        }
}
// END FILE "/VMap/src/com/vynance/modules/EnumClusters.as"

// BEGIN FILE "/VMap/src/com/vynance/modules/EnumExchanges.as"
package com.vynance.modules
{
        public class EnumExchanges
        {
            public static const AMEX:int = 0;
            public static const NYSE:int = 1;
            public static const NSDQ:int = 2;
            public static const COUNT:int = 3;
        }
}
// END FILE "/VMap/src/com/vynance/modules/EnumExchanges.as"

// BEGIN FILE "/VMap/src/com/vynance/modules/EnumMarketStatuses.as"
package com.vynance.modules
{
        public class EnumMarketStatuses
        {
            public static const PRE_OPEN:int = 0;
            public static const OPEN:int = 1;
            public static const CLOSED:int = 2;
        }
}
// END FILE "/VMap/src/com/vynance/modules/EnumMarketStatuses.as"

// BEGIN FILE "/VMap/src/com/vynance/modules/EnumTierTypes.as"
package com.vynance.modules
{
        public class EnumTierTypes
        {
            public static const None:int = 0;
            public static const Rows:int = 1;
            public static const Columns:int = 2;
        }
}
// END FILE "/VMap/src/com/vynance/modules/EnumTierTypes.as"

// BEGIN FILE "/VMap/src/com/vynance/utils/Signal.as"
package com.vynance.utils
{
        import com.vynance.model.Ticker;
        import com.vynance.modules.Constants;

        public class Signal
```

```
        {
                public static function compareSignal(value1:Number,
value2:Number):int
                {
                        var val1:int = Math.round(value1 * 100);
                        var val2:int = Math.round(value2 * 100);

                        if (val1 > val2)
                                return 1;
                        else if (val1 < val2)
                                return -1;
                        else
                                return 0;
                }

                public static function getColor(ticker:Ticker):int
                {
                        if (isNaN(ticker.signal))
                                return Constants.SIGNAL_COLOR_NA;
                        else if (compareSignal(Math.abs(ticker.signal), 5) >= 0)
                                return Constants.SIGNAL_COLOR_5_PLUS;
                        else if (compareSignal(Math.abs(ticker.signal), 4) >= 0)
                                return Constants.SIGNAL_COLOR_4_TO_5;
                        else if (compareSignal(Math.abs(ticker.signal), 3) >= 0)
                                return Constants.SIGNAL_COLOR_3_TO_4;
                        else if (compareSignal(Math.abs(ticker.signal), 2) >= 0)
                                return Constants.SIGNAL_COLOR_2_TO_3;
                        else if (compareSignal(Math.abs(ticker.signal), 1) >= 0)
                                return Constants.SIGNAL_COLOR_1_TO_2;
                        else if (compareSignal(Math.abs(ticker.signal), 0) >= 0)
                                return Constants.SIGNAL_COLOR_0_TO_1;
                        else
                                return Constants.SIGNAL_COLOR_NA;
                }
        }
}
// END FILE "/VMap/src/com/vynance/utils/Signal.as"

// BEGIN FILE "t.ASP"
<%
Response.Write(Time)
%>
// END FILE "t.ASP"
```

## Appendix B: Source Code for Signal

Below we give source code, written in R [R Project, 2017], for generating the `m.txt` and `s.txt` input files used by the front-end GUI source code in Appendix A. The entry function in this R code is `vm.univ(short.day = F)` (the input parameter is defined in a comment within this function). It uses a single file `mkt.data.txt` as an input (see Section 3). Together with the files `m.txt` and `s.txt`, this code also outputs another file `sig.delta.txt`, which contains (among other quantities – see Section 3) the last value of the signal computed by the R code. This file can also be used for debugging. The R code does not upload or download any files: `vm.univ()` is a call-back function for locally generating `m.txt` and `s.txt` instances.

```
vm.calc.signal <- function (vm.db)
{
      univ <- vm.db$tickers
      close <- vm.db$close
      high <- vm.db$high
      low <- vm.db$low
      last <- vm.db$last
      ind <- vm.db$ind.class
      cap <- vm.db$mkt.cap
      wt <- vm.db$weight

      n <- length(close)

      t <- vm.calc.t(vm.db)
      t <- rep(t, n)
      take <- close == 0
      t[take] <- 1

      x <- (1 - t) * close + t * (high + low) / 2
      ret <- rep(NA, n)
      y <- last / x
      take <- is.finite(y)
      ret[take] <- log(y[take]) * wt[take]
      ind.ret <- colSums(ret[take] * ind[take, ]) / colSums(ind[take, ])
      bad <- !is.finite(ind.ret)
      ind.ret[bad] <- 0
      ind.ret <- colSums(ind.ret * t(ind))
      ret <- ret - ind.ret
      mad.ret <- mad(ret[take])
      sig <- ret / mad.ret
      sig <- round(sig, 2)
      prev.sig <- vm.db$signal
      vm.db$delta <- round(abs(sig - vm.db$signal), 4)
      vm.db$signal <- sig
      take <- vm.db$bb == ""
      vm.db$delta[take] <- vm.db$signal[take] <- NA

      vm.db$delta.ix <- vm.sort.delta(vm.db$delta)
      vm.db$signal.ix <- vm.sort.signal(cap, vm.db$signal)
      vm.db$scrambled.sig <- vm.scramble(vm.db$signal)
}

vm.calc.stamp <- function (vm.db)
{
      if(vm.db$ssm >= vm.db$ssm.close)
            return(-1)

      if(vm.db$ssm <= vm.db$ssm.open)
            return(0)

      x <- vm.db$ssm - vm.db$ssm.open
      x <- ceiling(x / 60)
      return(x)
```

```
}

vm.calc.t <- function (vm.db)
{
      ssm <- vm.db$ssm

      if(ssm < vm.db$ssm.open)
            t <- 0
      else if(ssm > vm.db$ssm.close)
            t <- 1
      else
            t <- (ssm - vm.db$ssm.open) / (vm.db$ssm.close - vm.db$ssm.open)

      return(t)
}

vm.dump <- function ()
{
      x <- sapply(ls(pos = 1), function(x) storage.mode(get(x)))
      y <- names(x)[x == "function"]
      z <- grep("vm\\.", y)
      y <- y[z]
      save(list = y, file = "VM.RData")
      dump(y, "vm.code.txt")
}

vm.fix.sort <- function (x)
{
      take <- !is.finite(x)
      x[take] <- -9999
      return(x)
}

vm.map <- function (db.vm)
{
      convert.exch <- function(x)
      {
            x <- gsub("A", "0", x)
            x <- gsub("N", "1", x)
            x <- gsub("Q", "2", x)
            return(x)
      }

      tic <- db.vm$tickers
      ix <- db.vm$sector
      exch <- convert.exch(db.vm$exch)
      cap <- db.vm$mkt.cap
      cap.ix <- vm.sort.quant(tic, cap)
      liq <- db.vm$liquidity
      liq.ix <- vm.sort.quant(tic, liq)

      x <- cbind(
            tic,
            ix,
            exch,
            cap,
            cap.ix,
```

```
                liq,
                liq.ix)

        vm.write.table(x, db.vm$file.map)
}

vm.mkt.data <- function (vm.db)
{
        x <- read.delim(vm.db$file.data, header = T)
        x <- as.matrix(x)

        vm.db$tickers <- as.character(x[, "Ticker"])
        vm.db$sector <- as.numeric(x[, "Sector"])
        vm.db$exch <- as.character(x[, "Exchange"])
        vm.db$mkt.cap <- as.numeric(x[, "MktCap"])
        vm.db$liquidity <- as.numeric(x[, "Liquidity"])
        vm.db$close <- as.numeric(x[, "Close"])
        vm.db$last <- as.numeric(x[, "Last"])
        vm.db$high <- as.numeric(x[, "High"])
        vm.db$low <- as.numeric(x[, "Low"])
        vm.db$weight <- as.numeric(x[, "Weight"])
        bb <- as.character(x[, "IndNames"])
        bb[is.na(bb)] <- ""
        vm.db$bb <- bb
        vm.db$signal <- as.numeric(x[, "Signal"])

        ind.names <- unique(bb)
        n <- length(bb)
        k <- length(ind.names)
        ind <- matrix(0, n, k)
        for(i in 1:k)
                ind[, i] <- as.numeric(bb == ind.names[i])

        vm.db$ind.class <- ind
}

vm.scramble <- function (x)
{
        ix <- (1:length(x)) + 1
        y <- sin(sqrt(3) * ix + sqrt(7) * cos(sqrt(11) * ix))
        y <- round(y, 2)
        take <- y == 0
        z <- cos(sqrt(3) * ix + sqrt(7) * sin(sqrt(11) * ix))
        y[take] <- round(z[take], 2)
        x <- x * y
        return(x)
}

vm.sort.delta <- function (x)
{
        x <- as.numeric(x)
        x <- vm.fix.sort(x)
        n <- length(x)
        ix <- as.character(1:n)
        names(x) <- ix
        x <- sort(x)
```

```
        x[] <- ix
        ix <- x[ix]

        ix <- n - as.numeric(ix)
        return(ix)
}

vm.sort.quant <- function(tic, x)
{
        x <- vm.fix.sort(as.numeric(x))
        z <- tic
        ix <- as.character(1:length(x))
        names(z) <- names(x) <- ix
        x <- sort(x)

        z <- z[names(x)]

        take <- x == 0
        if(sum(take > 0))
        {
                y <- z[take]
                y <- sort(y)
                names(x)[take] <- names(y)
        }

        x[] <- ix
        ix <- x[ix]

        ix <- as.numeric(ix) - 1
        return(ix)
}

vm.sort.signal <- function(z, x)
{
        x <- vm.fix.sort(abs(x))
        z <- vm.fix.sort(z)

        n <- length(x)
        ix <- as.character(1:n)
        names(z) <- names(x) <- ix

        z <- sort(z)
        x <- x[names(z)]
        x <- sort(x)

        x[] <- ix
        ix <- x[ix]

        ix <- as.numeric(ix) - 1
        return(ix)
}

vm.ssm <- function(ssm = T)
{
        x <- as.integer(format(Sys.time(), "%H%M%S"))
        if(ssm)
        {
```

```
            s <- x - as.integer(x / 100) * 100
            x <- (x - s) / 100
            m <- x - as.integer(x / 100) * 100
            h <- (x - m) / 100
            x <- h * 3600 + m * 60 + s
        }
        return(x)
}

vm.univ <- function (short.day = F)
{
        vm.db <- new.env()

        vm.db$file.data <- "mkt.data.txt"
        vm.db$file.signal <- "s.txt"
        vm.db$file.map <- "m.txt"

        ### Max number of blinking tickers
        vm.db$max.deltas <- 25
        ### Market open: 9:30 AM
        vm.db$ssm.open <- 9.5 * 3600
        ### Market close: 4:00 PM (1:00 PM on short trading days)
        vm.db$ssm.close <- (16 - as.numeric(short.day) * 3) * 3600

        vm.mkt.data(vm.db)
        vm.map(vm.db)
        vm.write.signal(vm.db)
}

vm.write.signal <- function (vm.db)
{
        vm.db$ssm <- vm.ssm()
        x <- y <- vm.calc.stamp(vm.db)
        if(y > 0)
                vm.calc.signal(vm.db)

        sig.ix <- vm.db$signal.ix

        for(i in 1:length(vm.db$signal))
        {
                delta <- ""
                if(y == 0)
                {
                        sig <- "*"
                        ix <- i - 1
                        ix <- sig.ix[i]
                }
                else
                {
                        sig <- vm.db$scrambled.sig[i]
                        ix <- vm.db$signal.ix[i]
                        if(!is.na(vm.db$delta[i]) & vm.db$delta.ix[i] <
vm.db$max.deltas)
                                delta <- paste("\t", vm.db$delta.ix[i], sep = "")
                }
                x <- paste(x, ",", sig, "\t", ix, delta, sep = "")
        }
```

```
        vm.write.table(x, file = vm.db$file.signal)
        x <- cbind(vm.db$tickers, vm.db$scrambled.sig, vm.db$signal,
vm.db$signal.ix, vm.db$delta, vm.db$delta.ix)
        hdr <- c("Ticker", "Scrambled.Signal", "Signal", "Signal.ix", "Delta",
"Delta.ix")
        x <- rbind(hdr, x)
        vm.write.table(x, file = "sig.delta.txt", T)
}

vm.write.table <- function (x, file, last.return = F)
{
        if(last.return)
        {
                write.table(x, file = file, quote = F, row.names = F, col.names =
F, sep = "\t")
                return(1)
        }

        single.line <- F

        if(is.matrix(x))
                if(nrow(x) == 1)
                        single.line <- T

        if(is.vector(x))
                single.line <- T

        if(single.line)
        {
                write.table(x, file = file, quote = F, row.names = F, col.names =
F, sep = "\t", eol = "")
                return(1)
        }

        y <- x[nrow(x), ]
        x <- x[-nrow(x), ]
        z <- y[1]
        if(length(y) > 1)
                for(i in 2:length(y))
                        z <- paste(z, "\t", y[i], sep = "")
        write.table(x, file = file, quote = F, row.names = F, col.names = F,
sep = "\t")
        write.table(z, file = file, quote = F, row.names = F, col.names = F,
sep = "\t", eol = "", append = T)
}
```

## Appendix C: DISCLAIMERS

Wherever the context so requires, the masculine gender includes the feminine and/or neuter, and the singular form includes the plural and vice-versa. The author of this paper ("Author") and his affiliates including without limitation Quantigic® Solutions LLC ("Author's Affiliates" or "his Affiliates") make no implied or express warranties or any other representations whatsoever, including without limitation implied warranties of merchantability and fitness for a particular purpose, in connection with or with regard to the content of this paper, including without limitation any code or algorithms contained herein, and any supplementary files or data for which download links may be provided herein ("Content").

The reader may use the Content solely at his/her/its own risk and the reader shall have no claims whatsoever against the Author or his Affiliates and the Author and his Affiliates shall have no liability whatsoever to the reader or any third party whatsoever for any loss, expense, opportunity cost, damages or any other adverse effects whatsoever relating to or arising from the use of the Content by the reader including without any limitation whatsoever: any direct, indirect, incidental, special, consequential or any other damages incurred by the reader, however caused and under any theory of liability; any loss of profit (whether incurred directly or indirectly), any loss of goodwill or reputation, any loss of data suffered, cost of procurement of substitute goods or services, or any other tangible or intangible loss; any reliance placed by the reader on the completeness, accuracy or existence of the Content or any other effect of using the Content; and any and all other adversities or negative effects the reader might encounter in using the Content irrespective of whether the Author or his Affiliates is or are or should have been aware of such adversities or negative effects.

The source code included in Appendix A and Appendix B hereof is part of the copyrighted source code of Quantigic® Solutions LLC and is provided herein with the express permission of Quantigic® Solutions LLC. The copyright owner retains all rights, title and interest in and to its copyrighted source code included in Appendices A and B hereof and any and all copyrights therefor. Quantigic® and Vynance® are registered service marks of Quantigic® Solutions LLC.

## Supplementary Materials

The zipped folder `assets.zip` (see Appendix A) can be downloaded from

https://www.drivehq.com/file/DFPublishFile.aspx/FileID4837464317/Keytngjk5r5da7i/assets.zip

## Tables

| Absolute value of the Signal | Hex HTML color code used in the ActionScript source code (see Appendix A) | Color (approximate description) | Approximate color |
|---|---|---|---|
| N/A | 0xB4B4B4 | Lighter grey | 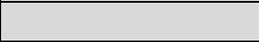 |
| Between 0 and 1 | 0x666666 | Darker grey | 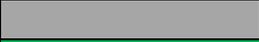 |
| Between 1 and 2 | 0x40B06C | Green | 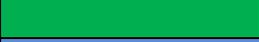 |
| Between 2 and 3 | 0x3380C2 | Blue | 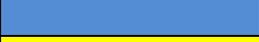 |
| Between 3 and 4 | 0xF4D701 | Yellow | 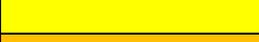 |
| Between 4 and 5 | 0xFF9C2C | Orange | 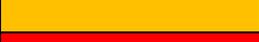 |
| Greater than 5 | 0xF53636 | Red | 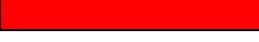 |

**Table 1.** Color-coding of the signal.

| Sector (Cluster) name | Numeric code used internally in the ActionScript source code (see Appendix A) |
|---|---|
| Cyclicals | 0 |
| Energy | 1 |
| Financials | 2 |
| Healthcare | 3 |
| Industrials | 4 |
| Materials | 5 |
| Non-Cyclicals | 6 |
| Technology | 7 |
| Telecom | 8 |
| Utilities | 9 |

**Table 2.** Sectors (Clusters) and their numeric codes used internally in the source code.

**Figures**

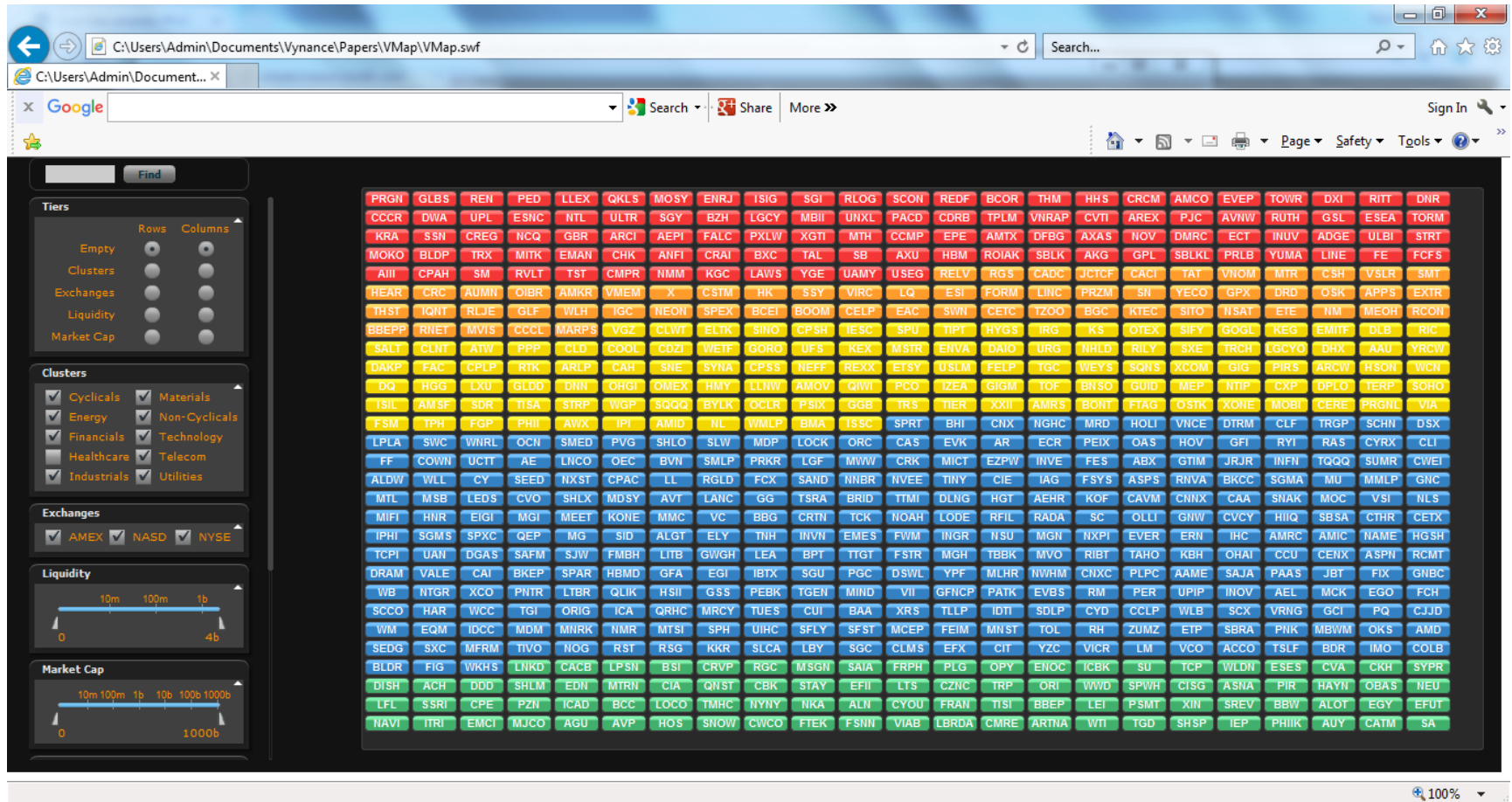

**Figure 1.** VMap appearance in a browser (Internet Explorer). The tickers are sorted according to the decreasing (absolute) value of the Signal (default). The data used is from April 29, 2016. The Signal is simulated to be computed at the market close (i.e., in Eq. (2): t = 1; Last = Close for April 29, 2016; High & Low are for April 29, 2016 at the close; and Close for April 28, 2016 does not contribute). The total number of tickers is 5,126. For simplicity, when computing the Signal, the Industries are identified with the Sectors.

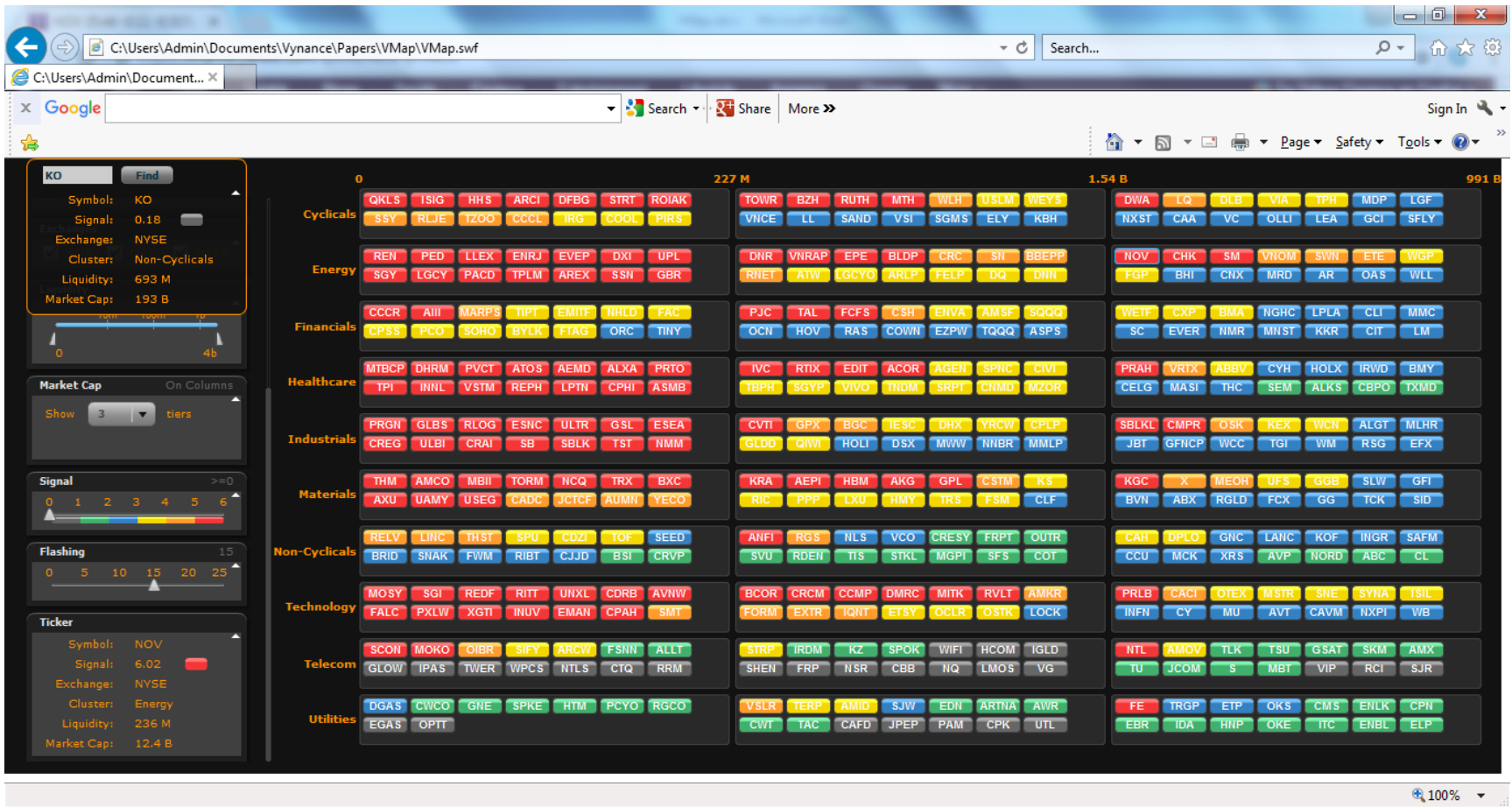

**Figure 2.** VMap appearance in a browser: same as in Figure 1, except this screenshot shows the lower part of the left panel (with the data for a moused-over ticker NOV displayed). In addition, it shows the data for the searched ticker KO (The Coca-Cola Co) in the upper left corner. As an illustration, Clusters are tiered on the rows, and Market Capitalization is tiered on the columns (with 3 tiers).